# A Bayesian Conjugate-Gradient Method

Jon Cockayne[*], Chris J. Oates[†], Ilse C.F. Ipsen[‡] and Mark Girolami[§]

December 18, 2018

A fundamental task in numerical computation is the solution of large linear systems. The conjugate gradient method is an iterative method which offers rapid convergence to the solution, particularly when an effective preconditioner is employed. However, for more challenging systems a substantial error can be present even after many iterations have been performed. The estimates obtained in this case are of little value unless further information can be provided about, for example, the magnitude of the error. In this paper we propose a novel statistical model for this error, set in a Bayesian framework. Our approach is a strict generalisation of the conjugate gradient method, which is recovered as the posterior mean for a particular choice of prior. The estimates obtained are analysed with Krylov subspace methods and a contraction result for the posterior is presented. The method is then analysed in a simulation study as well as being applied to a challenging problem in medical imaging.

## 1 Introduction

This paper presents an iterative method for solution of systems of linear equations of the form

$$A\boldsymbol{x}^* = \boldsymbol{b} \tag{1}$$

where $A \in \mathbb{R}^{d\times d}$ is an invertible matrix and $\boldsymbol{b} \in \mathbb{R}^d$ is a vector, each given, while $\boldsymbol{x}^* \in \mathbb{R}^d$ is to be determined. The principal novelty of our method, in contrast to existing approaches, is that its output is a *probability distribution* over vectors $\boldsymbol{x} \in \mathbb{R}^d$ which reflects knowledge about $\boldsymbol{x}^*$ after expending a limited amount of computational effort. This allows the output of the method to be used, in a principled *anytime* manner, tailored to reflect a constrained computational budget. In a special case, the mode of this distribution coincides with the estimate provided by the standard conjugate gradient method, whilst the probability mass is proven to contract onto $\boldsymbol{x}^*$ as more iterations are performed.

[*]University of Warwick (j.cockayne@warwick.ac.uk).
[†]Newcastle University and the Alan Turing Institute (chris.oates@ncl.ac.uk).
[‡]North Carolina State University (ipsen@ncsu.edu)
[§]Imperial College London and the Alan Turing Institute (m.girolami@imperial.ac.uk).

Challenging linear systems arise in a wide variety of applications; of these, partial differential equations (PDEs) should be emphasised, as these arise frequently throughout the applied sciences and in engineering [Evans, 2010]. Finite element and finite difference discretisations of PDEs each yield large, sparse linear systems which can sometimes be highly ill-conditioned, such as in the classically ill-posed backwards heat equation [Evans, 2010]. Even for linear PDEs, a detailed discretisation may be required. This can result in a linear system with billions of degrees of freedom and require specialised algorithms to be even approximately solved practically [e.g. Reinarz et al., 2018]. Another example arises in computation with Gaussian measures [Bogachev, 1998, Rasmussen, 2004], in which analytic covariance functions, such as the exponentiated quadratic, give rise to challenging linear systems. This has an impact in a number of related fields, such as symmetric collocation solution of PDEs [Fasshauer, 1999, Cockayne et al., 2016], numerical integration [Larkin, 1972, Briol et al., 2018] and generation of spatial random fields [Besag and Green, 1993, Parker and Fox, 2012, Schäfer et al., 2017]. In the latter case, large linear systems must often be solved to sample from these fields, such as in models of tropical ocean surface winds [Wikle et al., 2001] where systems may again be billion-dimensional. Thus, it is clear that there exist many important situations in which error in the solution of a linear system cannot practically be avoided.

## 1.1 Linear Solvers

The solution of linear systems is one of the most ubiquitous problems in numerical analysis and Krylov subspace methods [Hestenes and Stiefel, 1952, Liesen and Strakos, 2012] are among the most successful at obtaining an approximate solution at low cost. Krylov subspace methods belong to the class of *iterative methods* [Saad, 2003], which construct a sequence $(\boldsymbol{x}_m)$ that approaches $\boldsymbol{x}^*$ and can be computed in an efficient manner. Iterative methods provide an alternative to *direct methods* [Davis, 2006, Allaire and Kaber, 2008] such as the LU or Cholesky decomposition, which generally incur higher cost as termination of the algorithm after $m < d$ iterations is not meaningful. In certain cases an iterative method can produce an accurate approximation to $\boldsymbol{x}^*$ with reduced computational effort and memory usage compared to a direct method.

The conjugate gradient (CG) method [Hestenes and Stiefel, 1952] is a popular iterative method, and perhaps the first instance of a Krylov subspace method. The error arising from CG can be shown to decay exponentially in the number of iterations, but convergence is slowed when the system is poorly conditioned. As a result, there is interest in solving equivalent *preconditioned* systems [Allaire and Kaber, 2008], either by solving $P^{-1}A\boldsymbol{x}^* = P^{-1}\boldsymbol{b}$ (left-preconditioning) or $AP^{-1}P\boldsymbol{x}^* = \boldsymbol{b}$ (right-preconditioning), where $P$ is chosen both so that $P^{-1}A$ (or $AP^{-1}$) has a lower condition number than $A$ itself, and so that computing the solution of systems $P\boldsymbol{y} = \boldsymbol{c}$ is computationally inexpensive for arbitrary $\boldsymbol{y}$ and $\boldsymbol{c}$. Effective preconditioning can dramatically improve convergence of CG, and of Krylov subspace methods in general, and is recommended even for well-conditioned systems, owing to how rapidly conjugacy is lost in CG when implemented numerically. One reasonably generic method for sparse systems involves approximate factorisation of the matrix, through an incomplete LU or incomplete Cholesky decomposition [e.g. Ajiz and Jennings, 1984, Saad, 1994]. Other common approaches exploit the structure of the problem. For example, in numerical solution of PDEs a coarse discretisation of the system can be used to construct a preconditioner for a finer discretisation [e.g. Bramble et al., 1990]. A more detailed survey of preconditioning methods can be found in many standard texts, such as Benzi [2002] and Saad [2003]. However, no approach is universal, and in general careful analysis of the structure of the problem is required to determine an effective preconditioner [Saad, 2003, p. 283]. At worst, constructing a good preconditioner can be as difficult as solving the linear system itself.

In situations where numerical error cannot practically be made negligible, an estimate for the error $\boldsymbol{x}_m - \boldsymbol{x}^*$ must accompany the output $\boldsymbol{x}_m$ of any linear solver. The standard approach is to analytically bound $\|\boldsymbol{x}_m - \boldsymbol{x}^*\|$ by some function of the residual $\|A\boldsymbol{x}_m - \boldsymbol{b}\|$, for appropriate choices of norms, then to monitor the decay of the relative residual. In implementations, the algorithm is usually terminated when this reaches machine precision, which can require a very large number of iterations, and substantial computational effort. This often constitutes the principal bottleneck in contemporary applications. The contribution of this paper is to demonstrate how Bayesian analysis can be used to develop a richer, probabilistic description for the error in estimating the solution $\boldsymbol{x}^*$ with an iterative method. From a user's perspective, this means that solutions from the presented method can still be used in a principled way, even when only a small number of iterations can be afforded.

## 1.2 Probabilistic Numerical Methods

The concept of a probabilistic numerical method dates back to Larkin [1972]. The principal idea is that problems in numerical analysis can be cast as inference problems and are therefore amenable to statistical treatment. *Bayesian* probabilistic numerical methods [Cockayne et al., 2017] posit a prior distribution for the unknown, in our case $\boldsymbol{x}^*$, and condition on a finite amount of information about $\boldsymbol{x}^*$ to obtain a posterior that reflects the level of uncertainty in $\boldsymbol{x}^*$, given the finite information obtained. In contemporary applications, it is common for several numerical methods to be composed in a *pipeline* to perform a complex task. For example, climate models [such as Roeckner et al., 2003] involve large systems of coupled differential equations. To simulate from these models, many approximations must be combined. Bayesian probabilistic numerical methods are of particular interest in this setting, as a probabilistic description of error can be coherently propagated through the pipeline to describe the structure of the overall error and study the contribution of each component of the pipeline to that error [Hennig et al., 2015]. As many numerical methods rely on linear solvers, such as the CG method, understanding the error incurred by these numerical methods is critical. Other works to recently highlight the value of statistical thinking in this application area includes Calvetti et al. [2018].

In recent work, Hennig [2015] treated the problem of solving Eq. (1) as an infer-ence problem for the matrix $A^{-1}$, and established correspondence with existing iterative methods by selection of different matrix-valued Gaussian priors within a Bayesian framework. This approach was explored further in Bartels and Hennig [2016]. There, it was observed that the posterior distribution over the matrix in Hennig [2015] produces the same factors as in the LU or Cholesky decompositions[1]. Our contribution takes a fundamentally different approach, in that a prior is placed on the solution $\boldsymbol{x}^*$ rather than on the matrix $A^{-1}$. There are advantages to the approach of Hennig [2015], in that solution of multiple systems involving the same matrix is trivial. However we argue that it is more intuitive to place a prior on $\boldsymbol{x}^*$ than on $A^{-1}$, as one might more easily reason about the solution to a system than the elements of the inverse matrix. Furthermore, the approach of placing a prior on $\boldsymbol{x}^*$ is unchanged by any left-preconditioning of the system, while the prior of Hennig [2015] is not preconditioner-invariant.

**Contribution** The main contributions of this paper are as follows:

- The *Bayesian conjugate gradient* (BayesCG) method is proposed for solution of linear systems. This is a novel probabilistic numerical method in which both prior and posterior are defined on the solution space for the linear system, $\mathbb{R}^d$. We argue that placing a prior on the solution space is more intuitive than existing probabilistic numerical methods and corresponds more directly with classical iterative methods. This makes substitution of BayesCG for existing iterative solvers simpler for practitioners.

- The specification of the prior distribution is discussed in detail. Several natural prior covariance structures are introduced, motivated by preconditioners or Krylov subspace methods. In addition, a hierarchical prior is proposed in which all parameters can be marginalised, allowing automatic adjustment of the posterior to the scale of the problem. This discussion provides some generic prior choices to make application of BayesCG more straightforward for users unfamiliar with probabilistic numerical methods.

- It is shown that, for a particular choice of prior, the posterior mode of BayesCG coincides with the output of the standard CG method. An explicit algorithm is provided whose complexity is shown to be a small constant factor larger than that of the standard CG method. Thus, BayesCG can be efficiently implemented and could be used in place of classical iterative methods with marginal increase in computational cost.

- A thorough convergence analysis for the new method is presented, with computational performance in mind. It is shown that the posterior mean lies in a particular Krylov subspace, and rates of convergence for the mean and contraction for the posterior are presented. The distributional quantification of uncertainty provided by this method is shown to be conservative in general.

[1] Recall that the Cholesky decomposition is a symmetric version of the LU decomposition for symmetric positive-definite matrices.

The structure of the paper is as follows: In Section 2 BayesCG is presented and its inputs discussed. Its correspondence with CG is also established for a particular choice of prior. Section 3 demonstrates that the mean from BayesCG lies in a particular Krylov subspace and presents a convergence analysis of the method. In Section 4 the critical issue of prior choice is addressed. Several choices of prior covariance are discussed and a hierarchical prior is introduced to allow BayesCG to adapt to the scale of the problem. Section 5 contains implementation details, while in Section 6 the method is applied to a challenging problem in medical imaging which requires repeated solution of a linear system arising from the discretisation of a PDE. The paper concludes with a discussion in Section 7. Proofs of all theoretical results are provided in the electronic supplement.

# 2 Methods

We begin in Section 2.1 by defining a Bayesian probabilistic numerical method for the linear system in Eq. (1). In Section 2.2 a correspondence to the CG method is established. In Section 2.3 we discuss a particular choice of search directions that define BayesCG. Throughout this paper, note that $A$ is not required to be symmetric positive-definite, except for in Section 2.2.

## 2.1 Probabilistic Linear Solver

In this section we present a general probabilistic numerical method for solving Eq. (1). The approach taken is Bayesian, so that the method is defined by the choice of prior and the information on which the prior is to be conditioned. For this work, the information about $\boldsymbol{x}^*$ is linear and is provided by *search directions* $\boldsymbol{s}_i$, $i = 1, \ldots, m \ll d$, through the matrix-vector products

$$y_i := (\boldsymbol{s}_i^\top A)\boldsymbol{x}^* = \boldsymbol{s}_i^\top \boldsymbol{b}. \tag{2}$$

The matrix-vector products on the right-hand-side are assumed to be computed without error[2], which implies a likelihood model in the form of a Dirac distribution:

$$p(\boldsymbol{y}|\boldsymbol{x}) = \delta(\boldsymbol{y} - S_m^\top A\boldsymbol{x}). \tag{3}$$

This section assumes the search directions are given *a-priori*. The specific search directions which define BayesCG will be introduced in Section 2.3.

In general the recovery of $\boldsymbol{x}^*$ from $m < d$ pieces of information is ill-posed. The prior distribution serves to regularise the problem, in the spirit of Tikhonov [1963], Stuart [2010]. Linear information is well-adapted to inference with stable distributions[3] such as the Gaussian or Cauchy distributions, in that the posterior distribution is available in closed-form. Optimal estimation with linear information is also well-understood [cf. Traub et al., 1988]. To proceed, let $\boldsymbol{x}$ be a random variable, which will be used to

[2] i.e. in exact arithmetic

[3] Let $X_1$ and $X_2$ be independent copies of a random variable $X$. Then $X$ is said to be *stable* if, for any constants $\alpha, \beta > 0$, the random variable $\alpha X_1 + \beta X_2$ has the same distribution as $\gamma X + \delta$ for some constants $\gamma > 0$ and $\delta$.

model epistemic uncertainty regarding the true solution $\boldsymbol{x}^*$, and endow $\boldsymbol{x}$ with the prior distribution

$$p(\boldsymbol{x}) = \mathcal{N}(\boldsymbol{x}; \boldsymbol{x}_0, \Sigma_0) \tag{4}$$

where $\boldsymbol{x}_0$ and $\Sigma_0$ are each assumed to be known *a-priori*, an assumption that will be relaxed in Section 4. It will be assumed throughout that $\Sigma_0$ is a symmetric and positive-definite matrix.

Having specified the prior and the information, there exists a unique Bayesian probabilistic numerical method which outputs the conditional distribution $p(\boldsymbol{x}|\boldsymbol{y}_m)$ [Cockayne et al., 2017] where $\boldsymbol{y}_m = [y_1, \ldots, y_m]^\top$ satisfies $\boldsymbol{y}_m = S_m^\top A\boldsymbol{x}^* = S_m^\top \boldsymbol{b}$, and $S_m$ denotes the matrix whose columns are $\boldsymbol{s}_1, \ldots, \boldsymbol{s}_m$. This is made clear in the following result:

**Proposition 1** (Probabilistic Linear Solver)**.** *Let* $\Lambda_m = S_m^\top A \Sigma_0 A^\top S_m$ *and* $\boldsymbol{r}_0 = \boldsymbol{b} - A\boldsymbol{x}_0$. *Then the posterior distribution is given by*

$$p(\boldsymbol{x}|\boldsymbol{y}_m) = \mathcal{N}(\boldsymbol{x}; \boldsymbol{x}_m, \Sigma_m)$$

*where*

$$\boldsymbol{x}_m = \boldsymbol{x}_0 + \Sigma_0 A^\top S_m \Lambda_m^{-1} S_m^\top \boldsymbol{r}_0 \tag{5}$$

$$\Sigma_m = \Sigma_0 - \Sigma_0 A^\top S_m \Lambda_m^{-1} S_m^\top A \Sigma_0 \tag{6}$$

This provides a distribution on $\mathbb{R}^d$ that reflects the state of knowledge given the information contained in $\boldsymbol{y}_m$. The mean, $\boldsymbol{x}_m$, could be viewed as an approximation to $\boldsymbol{x}^*$ that might be provided by a numerical method. From a computational perspective, the presence of the $m \times m$ matrix $\Lambda_m^{-1}$ could be problematic, as this implies a second linear system must be solved, albeit at a lower cost $\mathcal{O}(m^3)$. This could be addressed to some extent by updating $\Lambda_m^{-1}$ iteratively using the Woodbury matrix inversion lemma, though this would not reduce the overall cost. However, as the search directions can be chosen arbitrarily, this motivates a choice which *diagonalises* $\Lambda_m$, to make the inverse trivial. This will be discussed further in Section 2.3.

Note that the posterior distribution is singular, in that $\det(\Sigma_m) = 0$. This is natural since what uncertainty remains in directions not yet explored is simply the restriction, in the measure-theoretic sense, of the prior to the subspace orthogonal to the columns of $S_m^\top A$. As a result, the posterior distribution is concentrated on a linear subspace of $\mathbb{R}^d$. Singularity of the posterior makes computing certain quantities difficult, such as posterior probabilities. Nevertheless, $\Sigma_m$ can be decomposed using techniques such as the singular-value decomposition, so sampling from the posterior is straightforward.

Define a generic inner-product of two vectors in $\mathbb{R}^d$ by $\langle \boldsymbol{x}, \boldsymbol{x}' \rangle_M = \boldsymbol{x}^\top M \boldsymbol{x}'$, with associated norm $\| \cdot \|_M$. Note that for this to define a norm, it is required that $M$ be a positive-definite matrix. The following basic result establishes that the posterior covariance provides a meaningful connection to the error of $\boldsymbol{x}_m$, when viewed as a point estimator:

**Proposition 2.**

$$\frac{\|\boldsymbol{x}_m - \boldsymbol{x}^*\|_{\Sigma_0^{-1}}}{\|\boldsymbol{x}_0 - \boldsymbol{x}^*\|_{\Sigma_0^{-1}}} \leq \sqrt{\operatorname{tr}(\Sigma_m \Sigma_0^{-1})}$$

Thus the right hand side provides an upper bound on the relative error of the estimator $\boldsymbol{x}_m$ in the $\Sigma_0^{-1}$-norm. This is a weak result and tighter results for specific search directions are provided later. In addition to bounding the error $\boldsymbol{x}_m - \boldsymbol{x}^*$ in terms of the posterior covariance $\Sigma_m$, we can also compute the rate of contraction of the posterior covariance itself:

**Proposition 3.**

$$\mathrm{tr}(\Sigma_m \Sigma_0^{-1}) = d - m$$

The combination of Propositions 2 and 3 implies that the posterior mean $\boldsymbol{x}_m$ is consistent and, since the posterior covariance characterises the width of the posterior, Proposition 3 can be viewed as a posterior contraction result. This result is intuitive; after exploring $m$ linearly independent search directions, $\boldsymbol{x}^*$ has been perfectly identified in an $m$-dimensional linear subspace of $\mathbb{R}^d$. Thus, after adjusting for the weighting of $\mathbb{R}^d$ provided by the prior covariance $\Sigma_0$, it is natural that an appropriate measure of the size of the posterior should also converge at a rate that is linear.

## 2.2 Correspondence with the Conjugate Gradient Method

In this section we examine the correspondence of the posterior mean $\boldsymbol{x}_m$ described in Proposition 1 with the CG method. It is frequently the case that Bayesian probabilistic numerical methods have some classical numerical method as their mean, due to the characterisation of the conditional mean of a probability distribution as the $L_2$-best element of the underlying space consistent with the information provided [Diaconis, 1988, Cockayne et al., 2017].

**The Conjugate Gradient Method** A large class of iterative methods for solving linear systems defined by positive-definite matrices $A$ can be motivated by sequentially solving the following minimisation problem:

$$\boldsymbol{x}_m = \arg\min_{\boldsymbol{x} \in \mathcal{K}_m} \|\boldsymbol{x} - \boldsymbol{x}^*\|_A$$

where $\mathcal{K}_m$ is a sequence of $m$-dimensional linear subspaces of $\mathbb{R}^d$. It is straightforward to show that this is equivalent to:

$$\boldsymbol{x}_m = \arg\min_{\boldsymbol{x} \in \mathcal{K}_m} f(\boldsymbol{x})$$

where $f(\boldsymbol{x}) = \frac{1}{2}\boldsymbol{x}^\top A \boldsymbol{x} - \boldsymbol{x}^\top \boldsymbol{b}$ is a convex quadratic functional. Let $S_m \in \mathbb{R}^{d \times m}$ denote a matrix whose columns are arbitrary linearly independent search directions $\boldsymbol{s}_1, \dots, \boldsymbol{s}_m$, with $\mathrm{range}(S_m) = \mathcal{K}_m$. Let $\boldsymbol{x}_0$ denote an arbitrary starting point for the algorithm. Then $\boldsymbol{x}_m = \boldsymbol{x}_0 + S_m \boldsymbol{c}$ for some $\boldsymbol{c} \in \mathbb{R}^m$ which can be computed by solving $\nabla f(\boldsymbol{x}_0 + S_m \boldsymbol{c}) = 0$. This yields:

$$\boldsymbol{x}_m = \boldsymbol{x}_0 + S_m (S_m^\top A S_m)^{-1} S_m^\top (\boldsymbol{b} - A\boldsymbol{x}_0) \tag{7}$$

In CG [Hestenes and Stiefel, 1952] the search directions are constructed to simplify the inversion in Eq. (7) by imposing that the search directions are $A$-conjugate, that is, $\langle \boldsymbol{s}_i^{\text{CG}}, \boldsymbol{s}_j^{\text{CG}} \rangle_A = 0$ whenever $i \neq j$. A set $\{\boldsymbol{s}_i\}$ of $A$-conjugate vectors is also said to be $A$-*orthogonal*, while if the vectors additionally have $\|\boldsymbol{s}_i\|_A = 1$ for each $i$ they are said to be $A$-*orthonormal*. For simplicity of notation, we will usually work with $A$-orthonormal search directions, but in most implementations of CG the normalisation step can introduce stability issues and is therefore avoided.

Supposing that such a set of $A$-orthonormal search directions can be found, Eq. (7) simplifies to

$$\boldsymbol{x}_m^{\text{CG}} = \boldsymbol{x}_0^{\text{CG}} + S_m^{\text{CG}} (S_m^{\text{CG}})^\top (\boldsymbol{b} - A\boldsymbol{x}_0^{\text{CG}}) \tag{8}$$

which lends itself to an iterative numerical method:

$$\boldsymbol{x}_m^{\text{CG}} = \boldsymbol{x}_{m-1}^{\text{CG}} + \boldsymbol{s}_m^{\text{CG}} (\boldsymbol{s}_m^{\text{CG}})^\top (\boldsymbol{b} - A\boldsymbol{x}_{m-1}^{\text{CG}}).$$

Search directions are also constructed iteratively, motivated by gradient descent on the function $f(\boldsymbol{x})$, whose negative gradient is given by $-\nabla f(\boldsymbol{x}) = \boldsymbol{b} - A\boldsymbol{x}$. The initial un-normalised search direction $\tilde{\boldsymbol{s}}_1^{\text{CG}}$ is chosen to be $\tilde{\boldsymbol{s}}_1^{\text{CG}} = \boldsymbol{r}_0^{\text{CG}} = \boldsymbol{b} - A\boldsymbol{x}_0^{\text{CG}}$, so that $\boldsymbol{s}_1^{\text{CG}} = \tilde{\boldsymbol{s}}_1^{\text{CG}} / \|\tilde{\boldsymbol{s}}_1^{\text{CG}}\|_A$. Letting $\boldsymbol{r}_m^{\text{CG}} = \boldsymbol{b} - A\boldsymbol{x}_m^{\text{CG}}$, subsequent search directions are given by

$$\tilde{\boldsymbol{s}}_m^{\text{CG}} := \boldsymbol{r}_{m-1}^{\text{CG}} - \langle \boldsymbol{s}_{m-1}^{\text{CG}}, \boldsymbol{r}_{m-1}^{\text{CG}} \rangle_A \boldsymbol{s}_{m-1}^{\text{CG}} \tag{9}$$

with $\boldsymbol{s}_m^{\text{CG}} = \tilde{\boldsymbol{s}}_m^{\text{CG}} / \|\tilde{\boldsymbol{s}}_m^{\text{CG}}\|_A$. This construction leads to search directions $\boldsymbol{s}_1^{\text{CG}}, \dots, \boldsymbol{s}_m^{\text{CG}}$ which form an $A$-orthonormal set.

Eq. 8 makes clear the following proposition, which shows that for a particular choice of prior the CG method is recovered as the posterior mean from Proposition 1:

**Proposition 4.** *Assume $A$ is symmetric and positive-definite. Let $\boldsymbol{x}_0 = \boldsymbol{0}$ and $\Sigma_0 = A^{-1}$. Then, taking $S_m = S_m^{CG}$, Eq.* (5) *reduces to $\boldsymbol{x}_m = \boldsymbol{x}_m^{CG}$.*

This result provides an intriguing perspective on the CG method, in that it represents the estimate produced by a rational Bayesian agent whose prior belief about $\boldsymbol{x}^*$ is modelled by $\boldsymbol{x} \sim \mathcal{N}(\boldsymbol{0}, A^{-1})$. Dependence of the prior on the inaccessible matrix inverse is in accordance with the findings in Hennig [2015] (Theorem 2.4 and Lemma 3.4), in which an analogous result was presented. As observed in that paper, the appearance of $A^{-1}$ in the prior covariance is not practically useful, as while the matrix inverse cancels in the expression for $\boldsymbol{x}_m$, it remains in the expression for $\Sigma_m$.

## 2.3 Search Directions

In this section the choice of search directions for the method in Proposition 1 will be discussed, initially by following an information-based complexity [Traub et al., 1988] argument. For efficiency purposes, a further consideration is that $\Lambda_m$ should be easy to invert. This naturally suggests that search directions should be chosen to be conjugate with respect to the matrix $A\Sigma_0 A^\top$, rather than $A$. Note that this approach *does not* require $A$ to be positive-definite, as $A\Sigma_0 A^\top$ is positive-definite for any non-singular $A$. Two choices of search direction will be discussed:

**Optimal Information** One choice is to formulate selection of $S_m$ in a decision-theoretic framework, to obtain *optimal information* in the nomenclature of Cockayne et al. [2017]. Abstractly, denote the probabilistic numerical method discussed above by $P[\cdot;\mu,S_m]:\mathbb{R}^d\to\mathcal{P}(\mathbb{R}^d)$, where $\mathcal{P}(\mathbb{R}^d)$ is the set of all distributions on $\mathbb{R}^d$. The function $P[\boldsymbol{b};\mu,S_m]$ takes a right-hand-side $\boldsymbol{b}\in\mathbb{R}^d$, together with a prior $\mu\in\mathcal{P}(\mathbb{R}^d)$ and a set of search directions $S_m$ and outputs the posterior distribution from Proposition 1. Thus $P[\boldsymbol{b};\mu,S_m]$ is a measure and $P[\boldsymbol{b};\mu,S_m](\mathrm{d}\boldsymbol{x})$ denotes its infinitesimal element.

For general $\mu\in\mathcal{P}(\mathbb{R}^d)$, define the *average risk* associated with the search directions $S_m$ to be

$$R(S_m,\mu)=\iint L(\boldsymbol{x},\boldsymbol{x}^*)P[A\boldsymbol{x}^*;\mu,S_m](\mathrm{d}\boldsymbol{x})\mu(\mathrm{d}\boldsymbol{x}^*) \tag{10}$$

where $L(\boldsymbol{x},\boldsymbol{x}^*)$ represents a loss incurred when $\boldsymbol{x}$ is used to estimate $\boldsymbol{x}^*$. This can be thought of as the performance of the probabilistic numerical method, averaged both over the class of problems described by $\mu$ and over the output of the method. *Optimal information* in this paper concerns selection of $S_m$ to minimise $R(S_m,\mu)$. The following proposition characterises optimal information for the posterior in Proposition 1 in the case of a squared-error loss function and when $\boldsymbol{x}_0=\boldsymbol{0}$. Let $A^{-\top}=(A^{-1})^\top$, and let $M^{\frac{1}{2}}$ denote a square-root of a symmetric positive-definite matrix $M$ with the property that $M^{\frac{\top}{2}}M^{\frac{1}{2}}=M$, where $M^{\frac{\top}{2}}=(M^{\frac{1}{2}})^\top$.

**Proposition 5.** *Suppose $\mu=\mathcal{N}(\boldsymbol{0},\Sigma_0)$ and consider the squared-error loss $L(\boldsymbol{x},\boldsymbol{x}^*)=\|\boldsymbol{x}-\boldsymbol{x}^*\|_M^2$ where $M$ is an arbitary symmetric positive-definite matrix. Optimal information for this loss is given by*

$$S_m=A^{-\top}M^{\frac{\top}{2}}\Phi_m$$

*where $\Phi_m$ is the matrix whose columns are the $m$ leading eigenvectors of $M^{\frac{1}{2}}\Sigma_0M^{\frac{\top}{2}}$, normalised such that $\Phi_m^\top\Phi_m=I$.*

The dependence of the optimal information on $A^{-\top}$ is problematic except for when $M=A^\top A$, which corresponds to measuring the performance of the algorithm through the residual $\|A\boldsymbol{x}_m-\boldsymbol{b}\|_2^2$. While this removes dependence on the inverse matrix, finding the search directions in this case requires computing the eigenvectors of $A\Sigma_0A^\top$, the complexity of which would dominate the cost of computing the posterior in Proposition 1.

**Conjugacy** A second, more practical method for obtaining search directions that diagonalise $\Lambda_m$ is similar to that taken in CG. Search directions are constructed which are conjugate to the matrix $A\Sigma_0A^\top$ by following a similar procedure to that described in Section 2.2.

**Proposition 6** (Conjugate Search Directions $\implies$ Iterative Method)**.** *Assume that the search directions are $A\Sigma_0A^\top$-orthonormal. Denote $\boldsymbol{r}_m=\boldsymbol{b}-A\boldsymbol{x}_m$. Then, $\boldsymbol{x}_m$ in Eq.* (5) *simplifies to*

$$\boldsymbol{x}_m=\boldsymbol{x}_{m-1}+\Sigma_0A^\top\boldsymbol{s}_m(\boldsymbol{s}_m^\top\boldsymbol{r}_{m-1})$$

*while to compute $\Sigma_m$ in Eq.* (6) *it suffices to store only the vectors $\Sigma_0A^\top\boldsymbol{s}_j$, for $j=1,\dots,m$.*

On the surface, the form of this posterior differs slightly from that in Proposition 1, in that the data are given by $\boldsymbol{s}_m^\top \boldsymbol{r}_{m-1}$ rather than $\boldsymbol{s}_m^\top \boldsymbol{r}_0$. However, when search directions are conjugate, the two expressions are equivalent:

$$\begin{aligned} \boldsymbol{s}_m^\top \boldsymbol{r}_{m-1} &= \boldsymbol{s}_m^\top \boldsymbol{b} - \boldsymbol{s}_m^\top A \boldsymbol{x}_{m-1} \\ &= \boldsymbol{s}_m^\top \boldsymbol{b} - \boldsymbol{s}_m^\top A \boldsymbol{x}_0 - \underbrace{\boldsymbol{s}_m^\top A \Sigma_0 A^\top S_{m-1}^\top}_{=0} \boldsymbol{r}_0 = \boldsymbol{s}_m^\top \boldsymbol{r}_0. \end{aligned} \tag{11}$$

Use of $\boldsymbol{s}_m^\top \boldsymbol{r}_{m-1}$ reduces the amount of storage required compared to direct application of Eq. (5). It also helps with stability as, while search directions can be shown to be conjugate mathematically, the accumulation of numerical error from floating point precision is such that numerical conjugacy may not hold, a point discussed further in Section 5.

An approach to constructing conjugate search directions for our probabilistic linear solver is now presented, again motivated by gradient descent.

**Proposition 7** (Bayesian Conjugate Gradient Method)**.** *Recall the definition of the residual* $\boldsymbol{r}_m = \boldsymbol{b} - A\boldsymbol{x}_m$*. Denote* $\tilde{\boldsymbol{s}}_1 = \boldsymbol{r}_0$ *and* $\boldsymbol{s}_1 = \tilde{\boldsymbol{s}}_1 / \|\tilde{\boldsymbol{s}}_1\|_{A\Sigma_0 A^\top}$*. For* $m > 1$ *let*

$$\tilde{\boldsymbol{s}}_m = \boldsymbol{r}_{m-1} - \langle \boldsymbol{s}_{m-1}, \boldsymbol{r}_{m-1} \rangle_{A\Sigma_0 A^\top} \boldsymbol{s}_{m-1}.$$

*Further, assume* $\tilde{\boldsymbol{s}}_m \neq \boldsymbol{0}$ *and let* $\boldsymbol{s}_m = \tilde{\boldsymbol{s}}_m / \|\tilde{\boldsymbol{s}}_m\|_{A\Sigma_0 A^\top}$*. Then for each* $m$*, the set* $\{\boldsymbol{s}_i\}_{i=1}^m$ *is* $A\Sigma_0 A^\top$*-orthonormal, and as a result* $\Lambda_m = I$*.*

This is termed a Bayesian *conjugate gradient* method for the same reason as in CG, as search directions are chosen to be the direction of gradient descent subject to a conjugacy requirement, albeit a different one than in standard CG. In the context of Proposition 4, note that the search directions obtained coincide with those obtained from CG when $A$ is symmetric positive-definite and $\Sigma_0 = A^{-1}$. Thus, BayesCG is a strict generalisation of CG. Note, however, that these search directions are constructed in a data-driven manner, in that they depend on the right-hand-side $\boldsymbol{b}$. This introduces a dependency on $\boldsymbol{x}^*$ through the relationship in Eq. 1 which is not taken into account in the conditioning procedure and leads to conservative uncertainty assessment, as will be demonstrated in Section 6.1.

# 3 BayesCG as a Krylov Subspace Method

In this section a thorough theoretical analysis of the posterior will be presented. Fundamental to the analysis in this section is the concept of a *Krylov subspace.*

**Definition 8** (Krylov Subspace)**.** *The* Krylov subspace $K_m(M, \boldsymbol{v})$*,* $M \in \mathbb{R}^{d\times d}$*,* $\boldsymbol{v} \in \mathbb{R}^d$ *is defined as*

$$K_m(M, \boldsymbol{v}) := \operatorname{span}(\boldsymbol{v}, M\boldsymbol{v}, M^2\boldsymbol{v}, \ldots, M^m\boldsymbol{v}).$$

*For a vector* $\boldsymbol{w} \in \mathbb{R}^d$*, the* shifted Krylov subspace *is defined as*

$$\boldsymbol{w} + K_m(M, \boldsymbol{v}) := \operatorname{span}(\boldsymbol{w} + \boldsymbol{v}, \boldsymbol{w} + M\boldsymbol{v}, \boldsymbol{w} + M^2\boldsymbol{v}, \ldots, \boldsymbol{w} + M^m\boldsymbol{v}).$$

It is well-known that CG is a Krylov subspace method for symmetric positive-definite matrices $A$ [Liesen and Strakos, 2012], meaning that

$$\boldsymbol{x}_m^{\mathrm{CG}} = \underset{\boldsymbol{x} \in \boldsymbol{x}_0 + K_{m-1}(A, \boldsymbol{r}_0)}{\arg\min} \|\boldsymbol{x} - \boldsymbol{x}^*\|_A.$$

It will now be shown that the posterior mean for BayesCG, presented in Proposition 6, is a Krylov subspace method. For convenience, let $K_m^* := \boldsymbol{x}_0 + K_m(\Sigma_0 A^\top A, \Sigma_0 A^\top \boldsymbol{r}_0)$.

**Proposition 9.** *The BayesCG mean $\boldsymbol{x}_m$ satisfies*

$$\boldsymbol{x}_m = \underset{\boldsymbol{x} \in K_{m-1}^*}{\arg\min} \|\boldsymbol{x} - \boldsymbol{x}^*\|_{\Sigma_0^{-1}}.$$

This proposition gives an alternate perspective on the observation that, when $A$ is symmetric positive-definite and $\Sigma_0 = A^{-1}$, the posterior mean from BayesCG coincides with $\boldsymbol{x}_m^{\mathrm{CG}}$: Indeed, for this choice of $\Sigma_0$, $K_m^*$ coincides with $\boldsymbol{x}_0 + K_m(A, \boldsymbol{r}_0)$ and furthermore, since under this choice of $\Sigma_0$ the norm minimised in Proposition 9 is $\|\cdot\|_A$, it is natural that the estimates $\boldsymbol{x}_m$ and $\boldsymbol{x}_m^{\mathrm{CG}}$ should be identical.

Proposition 9 allows us to establish a convergence rate for the BayesCG mean which is similar to that which can be demonstrated for CG. Let $\kappa(M) = \|M\|_2 \|M^{-1}\|_2$ denote the condition number of a matrix $M$ in the matrix 2-norm. Now, noting that $\kappa(\Sigma_0 A^\top A)$ his well-defined, as $\Sigma_0$ and $A$ are each nonsingular, we have:

**Proposition 10.**

$$\frac{\|\boldsymbol{x}_m - \boldsymbol{x}^*\|_{\Sigma_0^{-1}}}{\|\boldsymbol{x}_0 - \boldsymbol{x}^*\|_{\Sigma_0^{-1}}} \leq 2 \left( \frac{\sqrt{\kappa(\Sigma_0 A^\top A)} - 1}{\sqrt{\kappa(\Sigma_0 A^\top A)} + 1} \right)^m.$$

This rate is similar to the well-known convergence rate which for CG, in which $\kappa(\Sigma_0 A^\top A)$ is replaced by $\kappa(A)$. However, since it holds that $\kappa(A^\top A) \geq \kappa(A)$, the convergence rate for BayesCG will often be worse than that for CG, unless $\Sigma_0$ is chosen judiciously to reduce the condition number of $\kappa(\Sigma_0 A^\top A)$. Thus it appears that there is a price to be paid when uncertainty quantification is needed. This is unsurprising, as it is generally the case that uncertainty quantification is associated with additional cost over methods for which uncertainty quantification is not provided.

Nevertheless, the rate of convergence in Proposition 10 is significantly faster than the rate obtained in Proposition 2. The reason for this is that knowledge about how the search directions $S_m$ were chosen has been exploited. The directions used in BayesCG are motivated by gradient descent on $f(\boldsymbol{s})$. Thus, if gradient descent is an effective heuristic for the problem at hand, then the magnitude of the error $\boldsymbol{x}_m - \boldsymbol{x}^*$ will decrease at a rate which is sub-linear. The same cannot be said for $\mathrm{tr}(\Sigma_m \Sigma_0^{-1})$ which continues to converge linearly as proven in Proposition 3. Thus, the posterior covariance will in general be conservative when the BayesCG search directions are used. This is verified empirically in Section 6.1.

# 4 Prior Choice

The critical issue of prior choice is now examined. In Section 4.1 selection of the prior covariance structure will be discussed. Then in Section 4.2 a hierarchical prior will be introduced to address the scale of the prior.

## 4.1 Covariance Structure

When $A$ is symmetric positive-definite, one choice which has already been discussed is to set $\Sigma_0 = A^{-1}$, which results in a posterior mean equal to the output of CG. However correspondance of the posterior mean with CG does not in itself justify this modelling choice from a probabilistic perspective and moreover this choice is not practical, as access to $A^{-1}$ would give immediate access to the solution of Eq. Eq. (1). We therefore discuss some alternatives for the choice of $\Sigma_0$.

**Natural Prior** Taking inspiration from probabilistic numerical methods for PDEs [Cockayne et al., 2016, Owhadi, 2015], another natural choice presents itself: The object through which information about $\boldsymbol{x}^*$ is extracted is $\boldsymbol{b}$, so it is natural, and mathematically equivalent, to place a relatively uninformative prior on the elements of $\boldsymbol{b}$ rather than on $\boldsymbol{x}^*$ itself. If $\boldsymbol{b} \sim \mathcal{N}(\boldsymbol{0}, I)$ then the implied prior model for $\boldsymbol{x}^*$ is $\boldsymbol{x} \sim \mathcal{N}(\boldsymbol{0}, (A^\top A)^{-1})$. This prior is as impractical as that which aligns the posterior mean with CG, but has the attractive property that convergence is instantaneous when the search directions from Proposition 7 are used. To see this, observe that

$$
\begin{aligned}
\boldsymbol{s}_1 &= \frac{\boldsymbol{r}_0}{\|\boldsymbol{r}_0\|_{A\Sigma_0 A^\top}} = \frac{\boldsymbol{r}_0}{\|\boldsymbol{r}_0\|_2} && \text{(since } A\Sigma_0 A = I\text{)} \\
\implies \boldsymbol{x}_1 &= \boldsymbol{x}_0 + \frac{(A^\top A)^{-1} A^\top \boldsymbol{r}_0 (\boldsymbol{r}_0^\top \boldsymbol{r}_0)}{\|\boldsymbol{r}_0\|_2^2} && \text{(Proposition 6)} \\
&= \boldsymbol{x}_0 + A^{-1}(\boldsymbol{b} - A\boldsymbol{x}_0) = \boldsymbol{x}^* && \text{(since } \boldsymbol{r}_0 = \boldsymbol{b} - A\boldsymbol{x}_0\text{)}.
\end{aligned}
$$

Thus this prior is natural, in that when using the search directions from Proposition 7, convergence occurs in one iteration.

**Preconditioner Prior** For systems in which a preconditioner is available, the preconditioner can be thought of as providing an approximation to the linear operator $A$. Inspired by the impractical natural covariance $(A^\top A)^{-1}$, one approach proposed in this paper is to set $\Sigma_0 = (P^\top P)^{-1}$, when a preconditioner $P$ can be found. Since by design the action of $P^{-1}$ can be computed efficiently, so too can the action of $\Sigma_0$. As mentioned in Section 1.1, the availability of a good preconditioner is problem-dependent.

**Krylov Subspace Prior** The analysis presented in Section 3 suggests another potential prior, in which probability mass is distributed according to an appropriate Krylov subspace $K_n(M, \boldsymbol{b})$. Consider a distribution constructed as the linear combination

$$\boldsymbol{x}_K = \sum_{i=0}^{n} w_i M^i \boldsymbol{b} \tag{12}$$

where $\boldsymbol{w} := (w_0, \ldots, w_n) \sim \mathcal{N}(\mathbf{0}, \Phi)$ for some positive-definite matrix $\Phi$. The distribution on $\boldsymbol{x}_K$ induced by Eq. (12) is clearly Gaussian with mean $\mathbf{0}$. To determine its covariance, note that the above expression can be rewritten as $\boldsymbol{x}_K = K_n \boldsymbol{w}$, where $K_n \in \mathbb{R}^{d \times (n+1)}$ is the matrix whose columns form a basis of the Krylov subspace $K_n(M, \boldsymbol{b})$. A convenient choice is $K_n = [\boldsymbol{k}_0, \ldots, \boldsymbol{k}_n]$, where

$$\tilde{\boldsymbol{k}}_i = A^i \boldsymbol{b} - \sum_{j=0}^{i-1} \boldsymbol{b}^\top A^{(i+j)} \boldsymbol{b} \cdot A^j \boldsymbol{b}$$

and $\boldsymbol{k}_i = \tilde{\boldsymbol{k}}_i / \|\tilde{\boldsymbol{k}}_i\|_2$ Irrespective of choice of $K_n$, however, the covariance of $\boldsymbol{x}_K$ is given by $\mathbb{E}(\boldsymbol{x}_K \boldsymbol{x}_K^\top) = K_n \Phi K_n^\top$ so that $\boldsymbol{x}_K \sim \mathcal{N}(\mathbf{0}, K_n \Phi K_n^\top)$. One issue with this approach is that the computation of the matrix $K_n$ is of the same computational complexity as $n$ iterations of BayesCG, requiring $n$ matrix-vector products. To ensure that this cost does not dominate the procedure, it is necessary to take $n < m \ll d$. However, in this situation $\boldsymbol{x}^* \notin K_n(\boldsymbol{b}, M)$, so it is necessary to add additional probability mass on the space orthogonal to $K_n(M, \boldsymbol{b})$, to ensure that $\boldsymbol{x}^*$ lies in the prior support. To this end, let $K_n^\perp(\boldsymbol{b}, M) = \mathbb{R}^d \setminus K_n(\boldsymbol{b}, M)$, and let $K_n^\perp$ denote a matrix whose columns span $K_n^\perp(\boldsymbol{b}, M)$. Let $\boldsymbol{x}_K^\perp = K_n^\perp \boldsymbol{w}^\perp$, where $\boldsymbol{w}^\perp \sim \mathcal{N}(\mathbf{0}, \varphi I)$ for a scaling parameter $\varphi \in \mathbb{R}$. Then, the proposed Krylov subspace prior is given by

$$\boldsymbol{x} \; (= \boldsymbol{x}_0 + \boldsymbol{x}_K + \boldsymbol{x}_K^\perp) \; \sim \mathcal{N}\left(\boldsymbol{x}_0, K_n \Phi K_n^\top + \varphi K_n^\perp (K_n^\perp)^\top\right).$$

Practical issues associated with this approach are now discussed:

- **Choice of** $M$: It seems natural to place mass on the Krylov subspace which $\boldsymbol{x}_m$ occupies, that is $K_m(\Sigma_0 A^\top A, \Sigma_0 A^\top \boldsymbol{r}_0)$, but dependence of this subspace on the prior covariance that is being computed makes this choice circular. An alternative choice is to choose $M$ so that the projection of $\boldsymbol{x}^*$ into $K_m(M, \boldsymbol{b})$ converges rapidly in $m$ to the truth. The choice $M = A$ seems natural, due to the known rapid convergence of CG, and corresponds to a prior encoding of the intuition that "CG search directions tend to work well". Another option would be to take $M = P^{-1}A$ for some preconditioner $P$, but this final choice was not explored.

- **Selection of** $\Phi$ **and** $\varphi$: When $M = A$, the theoretical bound for the relative error of CG can be used to choose the elements of $\Phi$, given in Proposition 10 when $\Sigma_0 = A^{-1}$. Let $\xi < 1$. Then we propose to choose $\Phi$ to be a diagonal matrix, with diagonal entries $\Phi_{ii} = \left[2\sigma\xi^i\right]^2$ where $\sigma \in \mathbb{R}$ is a scale parameter. Based upon Proposition 10, the ideal choices for $\xi$ and $\sigma$ are $\xi = \frac{\kappa(A)-1}{\kappa(A)+1}$ and $\sigma = \|\boldsymbol{x}^*\|_A$. However, since these two quantities are not typically known *a priori*, estimates must in practice be used. The remaining parameter, $\varphi$, determines how much weight is given to the orthogonal component. Given the choice of $\Phi$, it is natural to choose $\varphi < [2\sigma\xi^{i+2}]^2$; this encodes the user's prior belief about how many iterations of standard CG are "typically needed".

- **Computation of** $K_n^\perp$: Computation of the complement of $K_n$ is equivalent to finding a basis of the set of all vectors $\boldsymbol{v}$ for which $K_n^\top \boldsymbol{v} = \boldsymbol{0}$; that is, computing the null-space of $K_n^\top$. This can be accomplished by QR decomposition. Recall that a QR-decomposition produces matrices $Q \in \mathbb{R}^{d\times d}$ and $R \in \mathbb{R}^{d\times n}$ such that $K_n^\top = QR$. The matrix $Q$ can be used to determine the null space of $K_n^\top$. After partitioning the matrix $Q$ as $Q = [Q_1, Q_2]$, where $Q_1 \in \mathbb{R}^{d\times(n+1)}$ and $Q_2 \in \mathbb{R}^{d\times(d-n-1)}$ are respectively the first $n+1$ and last $d-n-1$ columns of $Q$, it holds that $Q_2$ is an orthonormal matrix whose columns form a basis of the required null-space.

## 4.2 Covariance Scale

For the distributional output of BayesCG to be useful it must be well-calibrated. Loosely speaking, this means that the true solution $\boldsymbol{x}^*$ should typically lie in a region where most of the posterior probability mass is situated. As such, the scale of the posterior variance should have the ability to adapt and reflect the difficulty of the linear system at hand. This can be challenging, partially because the magnitude of the solution vector is *a-priori* unknown and partially because of the aforementioned fact that the dependence of $S_m$ on $\boldsymbol{x}^*$ is not accounted for in BayesCG.

In this section we propose to treat the prior scale as an additional parameter to be learned; that is we consider the prior model

$$p(\boldsymbol{x}|\nu) = \mathcal{N}(\boldsymbol{x}_0, \nu\Sigma_0)$$

where $\boldsymbol{x}_0, \Sigma_0$ are as before, while $\nu \in \mathbb{R}^+$. This can be viewed as a generalised version of the prior in Eq. (4), which is recovered when $\nu = 1$. In this section we consider learning $\nu$ in a hierarchical Bayesian framework, but we note that $\nu$ could also be heuristically calibrated. An example of such a heuristic procedure is outlined in Section S3.

The approach pursued below follows a standard approach in Bayesian linear regression [Gelman et al., 2014]. More generally, one could treat the entire covariance as unknown and perform similar conjugate analysis with an inverse-Wishart prior, though this extension was not explored. Consider then endowing $\nu$ with Jeffreys' (improper) reference prior:

$$p(\nu) \propto \nu^{-1}.$$

The conjugacy of this prior with the Gaussian distribution is such that the posterior marginal distributions $p(\nu|\boldsymbol{y}_m)$ and $p(\boldsymbol{x}|\boldsymbol{y}_m)$ can be found analytically. For the following proposition, $\mathsf{IG}$ denotes an inverse-gamma distribution, while $\mathsf{MVT}_m$ denotes a multivariate $t$ distribution with $m$ degrees of freedom.

**Proposition 11** (Hierarchical BayesCG)**.** *When $p(\boldsymbol{x}|\nu)$ and $p(\nu)$ are as above, the posterior marginal for $\nu$ is given by*

$$p(\nu|\boldsymbol{y}_m) = \mathsf{IG}\left(\frac{m}{2}, \frac{1}{2}\boldsymbol{r}_0^\top S_m \Lambda_m^{-1} S_m^\top \boldsymbol{r}_0\right)$$

*while the posterior marginal for $\boldsymbol{x}$ is given by*

$$p(\boldsymbol{x}|\boldsymbol{y}_m) = \mathsf{MVT}_m\left(\boldsymbol{x}_m, \frac{\boldsymbol{r}_0^\top S_m \Lambda_m^{-1} S_m^\top \boldsymbol{r}_0}{m}\Sigma_m\right).$$

*When the search directions are $A\Sigma_0 A^\top$-orthonormal, this simplifies to*

$$\begin{aligned} p(\nu|\boldsymbol{y}_m) &= \mathsf{IG}\left(\frac{m}{2}, \frac{m}{2}\nu_m\right) \\ p(\boldsymbol{x}|\boldsymbol{y}_m) &= \mathsf{MVT}_m\left(\boldsymbol{x}_m, \nu_m \Sigma_m\right) \end{aligned}$$

*where $\nu_m := \|S_m^\top \boldsymbol{r}_0\|_2^2/m$.*

Since $\boldsymbol{r}_0$ reflects the initial error $\boldsymbol{x}_0 - \boldsymbol{x}^*$, the quantity $\nu_m$ can be thought of as describing the difficulty of the problem. Thus in this approach the scale of the posterior is data-dependent.

# 5 Implementation

In this section some important details of the implementation of BayesCG are discussed.

**Numerical Breakdown of Conjugacy** In standard CG, as well as for the sequentially-computed search directions from Proposition 7, while the search directions are conjugate in exact arithmetic, the propagation of floating point error is such that *numerical* conjugacy can fail to hold. This is due to the iterative way in which the method is implemented. The implication is that, while mathematical convergence is guaranteed in $d$ iterations, $m > d$ iterations may be required in practice for both the CG estimate to converge to $\boldsymbol{x}^*$ and the BayesCG posterior to contract around it. When used in practise, CG is only run for $m \ll d$ iterations to mitigate the impact of conjugacy breakdown. This further highlights the importance of preconditioners, to ensure rapid convergence.

This phenomenon can be mitigated to some extent by the modification described in Eq. (11) which ensures that "local" conjugacy is exploited [Meurant, 2006]. While the new direction may not be numerically conjugate to $\boldsymbol{r}_0$, it is likely to be numerically conjugate to $\boldsymbol{r}_{m-1}$ by the construction in Proposition 7. Thus, computing the quantity $\boldsymbol{s}_m^\top \boldsymbol{r}_{m-1}$ rather than $\boldsymbol{s}_m^\top \boldsymbol{r}_0$ promotes continued stability and convergence of the posterior mean $\boldsymbol{x}_m$.

The impact on the posterior covariance also deserves comment. In the regime when $d > m$, $\sigma_m$ takes complex values so the posterior covariance will no longer be meaningful. The underlying issue is that the fact that $\Lambda_m \neq I$ when the search directions are not perfectly conjugate, and so the simplification exploited in Proposition 6 induces *overconfidence* in the resulting posterior. Since the same issue arises in search directions obtained from CG, we note that the work of Hennig [2015], which also exploits conjugacy, is likely to suffer from the same deficiency. However a full treatment of floating point error in the context of BayesCG is rather technical and is deferred for future work. In this direction inspiration may be taken from Parker and Fox [2012], where a similar analysis was performed.

To circumvent these issues for the purposes of our experiments we introduce the *batch-computed search directions* obtained by a Gram-Schmidt orthogonalisation procedure[4]:

$$\begin{aligned}\tilde{\boldsymbol{s}}_m^C &:= \boldsymbol{r}_{m-1} - \sum_{i=1}^{m-1} \left\langle \boldsymbol{s}_i^C, \boldsymbol{r}_{m-1} \right\rangle_{A\Sigma_0 A^\top} \boldsymbol{s}_{m-1}^C \\ \boldsymbol{s}_m^C &:= \tilde{\boldsymbol{s}}_m^C / \|\tilde{\boldsymbol{s}}_m^C\|_{A\Sigma_0 A^\top}.\end{aligned}$$

The batch-computed search directions are mathematically identical to the BayesCG search directions $\{\boldsymbol{s}_i\}_{i=1}^m$. However, by explicitly orthogonalising with respect to all $m-1$ previous directions, this batch procedure ensures that numerical conjugacy is maintained.

**Computational Cost** The cost of BayesCG is a constant factor higher than the cost of CG as three, rather than one, matrix-vector multiplications are required. Thus, the overall cost is $\mathcal{O}(md^2)$ when the search directions from Proposition 7 are used. When the batch-computed search directions are used an additional loop of complexity $\mathcal{O}(m)$ must be performed. Thus, the cost of the BayesCG algorithm with batch-computed search directions is $\mathcal{O}(m^2 d^2)$. Note that each of these costs assumes that $A$ and $\Sigma_0$ are dense marices; in the case of sparse matrices the cost of the matrix-vector multiplications is driven by the number of nonzero entries of each matrix rather than the dimension $d$.

**Termination Criteria** An appealing use of the posterior distribution might be to derive a probabilistic termination criterion for BayesCG. Recall from Proposition 2 that $\boldsymbol{x}_m$ approaches $\boldsymbol{x}^*$ at a rate bounded by $\sigma_m := \sqrt{\operatorname{tr}(\Sigma_m \Sigma_0^{-1})}$, and from Proposition 3 that $\operatorname{tr}(\Sigma_m \Sigma_0^{-1}) = d - m$. To decide in practice how many iterations of BayesCG should be performed we propose a termination criterion based upon the posterior distribution from Proposition 11:

$$\sigma_m^2 := \operatorname{tr}(\Sigma_m \Sigma_0^{-1}) \times \nu_m = (d-m)\nu_m$$

Thus, termination when $\sigma_m < \epsilon$, for some tolerance $\epsilon > 0$ that is user-specified, might be a useful criterion. However, Proposition 2 is extremely conservative, and since Proposition 10 establishes a much faster rate of convergence for $\|\boldsymbol{x}_m - \boldsymbol{x}^*\|_{\Sigma_0^{-1}}$ in the case of BayesCG search directions, this is likely to be an overcautious stopping criterion in the case of BayesCG. Furthermore, since this involves a data-driven estimate of scale,

[4]See [Giraud et al., 2005] for a numerically stable alternative.

**Algorithm 1** Computation of the posterior distribution described in Proposition 6. The implementation is optimised compared to that given in Proposition 6; see Supplement S2 for detail. Further note that, for clarity, all required matrix-vector multiplications have been left explicit, but for efficiency these should be calculated once-per-loop and stored. $\Sigma_m$ can be computed from this output as $\Sigma_m = \Sigma_0 - \Sigma_F \Sigma_F^\top$.

**procedure** BayesCG($A, \boldsymbol{b}, \boldsymbol{x}_0, \Sigma_0, \epsilon, m_{\max}$) ▷ ($\epsilon$ the tolerance)
  $\Sigma_F$ initialised to a matrix of size $(d \times 0)$ ▷ ($m_{\min}$ the minimum # iterations)
  $\boldsymbol{r}_0 \leftarrow \boldsymbol{b} - A\boldsymbol{x}_0$ ▷ ($m_{\max}$ the maximum # iterations)
  $\tilde{\boldsymbol{s}}_1 \leftarrow \boldsymbol{r}_0$
  $\tilde{\nu}_0 \leftarrow 0$
  **for** $m = 1, \dots, m_{\max}$ **do**
    $E^2 \leftarrow \tilde{\boldsymbol{s}}_m^\top A \Sigma_0 A^\top \tilde{\boldsymbol{s}}_m$
    $\alpha_m \leftarrow \frac{\boldsymbol{r}_{m-1}^\top \boldsymbol{r}_{m-1}}{E^2}$
    $\boldsymbol{x}_m \leftarrow \boldsymbol{x}_{m-1} + \alpha_m \Sigma_0 A^\top \tilde{\boldsymbol{s}}_m$
    $\boldsymbol{r}_m \leftarrow \boldsymbol{r}_{m-1} - A\boldsymbol{x}_m$
    $\Sigma_F \leftarrow [\Sigma_F, \Sigma_0 A^\top \tilde{\boldsymbol{s}}_m / E]$
    $\tilde{\nu}_m \leftarrow \tilde{\nu}_{m-1} + \frac{(\boldsymbol{r}_{m-1}^\top \boldsymbol{r}_{m-1})^2}{E^2}$
    **if** $\|r_m\|_2 < \epsilon$ **then**
      **break**
    **end if**
    $\beta_m \leftarrow \frac{\boldsymbol{r}_m^\top \boldsymbol{r}_m}{\boldsymbol{r}_{m-1}^\top \boldsymbol{r}_{m-1}}$
    $\tilde{\boldsymbol{s}}_{m+1} \leftarrow \boldsymbol{r}_m + \beta_m \tilde{\boldsymbol{s}}_m$
  **end for**
  $\nu_m \leftarrow \tilde{\nu}_m / m$
  **return** $\boldsymbol{x}_m, \Sigma_F, \nu_m$
**end procedure**

the term $\nu_m$ is not uniformly decreasing with $m$. As a result, in practise we advocate using a more traditional termination criterion based upon monitoring the residual; see Golub and Van Loan [2013, Section 11.3.8] for more detail. Further research is needed to establish whether the posterior distribution can provide a useful termination criterion.

Full pseudocode for the BayesCG method, including the termination criterion, is presented in Algorithm 1. Two algebraic simplifications have been exploited here relative to the presentation in the main text; these are described in detail in Section S2 of the supplement. A Python implementation can be found at github.com/jcockayne/bcg.

# 6 Numerical Results

In this section two numerical studies are presented. First we present a simulation study in which theoretical results are verified. Second we present an application to electrical impedance tomography, a challenging medical imaging technique in which linear systems must be repeatedly solved.

## 6.1 Simulation Study

The first experiment in this section is a simulation study, the goals of which are to empirically examine the convergence properties of BayesCG and to compare the output of the algorithm against the probabilistic approach of Hennig [2015].

For our simulation study, a matrix $A$ was generated by randomly drawing its eigenvalues $\lambda_1, \ldots, \lambda_d$ from an exponential distribution with parameter $\gamma$. A sparse, symmetric-positive definite matrix with these eigenvalues was then drawn using the MATLAB function `sprandsym`. The proportion of non-zero entries was taken to be 20%. Subsequently, a vector $\boldsymbol{x}^*$ was drawn from a reference distribution $\mu_{\text{ref}}$ on $\mathbb{R}^d$, and $\boldsymbol{b}$ was computed as $\boldsymbol{b} = A\boldsymbol{x}^*$. Throughout, the reference distribution for $\boldsymbol{x}^*$ was taken to be $\mu_{\text{ref}} = \mathcal{N}(\mathbf{0}, I)$. For this experiment $d = 100$ and $\gamma = 10$. In all cases the prior mean was taken to be $\boldsymbol{x}_0 = \mathbf{0}$. The prior covariance was alternately taken to be $\Sigma_0 = I$, $\Sigma_0 = A^{-1}$ and $\Sigma_0 = (P^\top P)^{-1}$ where $P$ was a preconditioner found by computing an incomplete Cholesky decomposition with zero fill-in. This decomposition is simply a Cholesky decomposition in which the (approximate) factor $\hat{L}$ has the same sparsity structure as $A$. The preconditioner is then given by $P = \hat{L}\hat{L}^\top$. The matrix $\hat{L}$ can be computed at a computational cost of $\mathcal{O}(\text{nnz}(A)^3)$ where $\text{nnz}(A)$ is the number of nonzero entries of $A$. Furthermore, $P^{-1}$ is cheap to apply because its Cholesky factor is explicit. In addition, the Krylov subspace prior introduced in Section 4.1 has been examined. While it has been noted that the choice $\Sigma_0 = A^{-1}$ is generally impractical, for this illustrative example $A^{-1}$ has been computed directly. Additional experimental results which apply the methodology discussed in this section to higher-dimensional problems is presented in Section S4.

**Point Estimation** In Figure 1 the convergence of the posterior mean $\boldsymbol{x}_m$ from BayesCG is contrasted with that of the output of CG, for many test problems $\boldsymbol{x}^*$ with a fixed sparse matrix $A$. As expected from the result of Proposition 9, the convergence of the BayesCG mean vector when $\Sigma_0 = I$ is slower than in CG. In this case, the speed of convergence for BayesCG is gated by $\kappa(A^\top A)$ which is larger than the corresponding $\kappa(A)$ for CG. The *a priori* optimal search directions also appear to yield a slower rate than the BayesCG search directions, owing to the fact that they do not exploit knowledge of $\boldsymbol{b}$. Similarly as expected, the posterior mean when $\Sigma_0 = A^{-1}$ is identical to the estimate for $\boldsymbol{x}_m$ obtained from CG. The fastest rate of convergence was achieved when $\Sigma_0 = (P^\top P)^{-1}$, which provides a strong motivation for using a preconditioner prior if such a preconditioner can be computed, though note that a preconditioned CG method would converge at a yet faster rate gated by $\kappa(P^{-1}A)$.

In the lower row of Figure 1 the convergence is shown when using batch-computed directions. Here convergence appears to be faster than when using the sequentially-computed directions, at correspondingly higher computational cost. The batch-computed directions provide an exact solution after $m = d$ iterations, in contrast to the sequentially-computed directions, for which numerical conjugacy may not hold.

Convergence for the Krylov subspace prior introduced in Section 4.1 is plotted in the right-hand column. The size of the computed subspace was set to $n = 20$, with $\sigma =$ $\|\boldsymbol{x}^*\|_A$ and $\xi = \frac{\kappa(A)-1}{\kappa(A)+1}$, as these quantites are easily computable in this simplified setting. The remaining parameter was set to $\gamma = 0.01$, to ensure that low prior weight was given to the remaining subspaces. With the sequentially computed directions significant numerical instability is observed starting at $m = 20$. This does not occur with the batch computed directions, where a jump in the convergence rate is seen at this iteration.

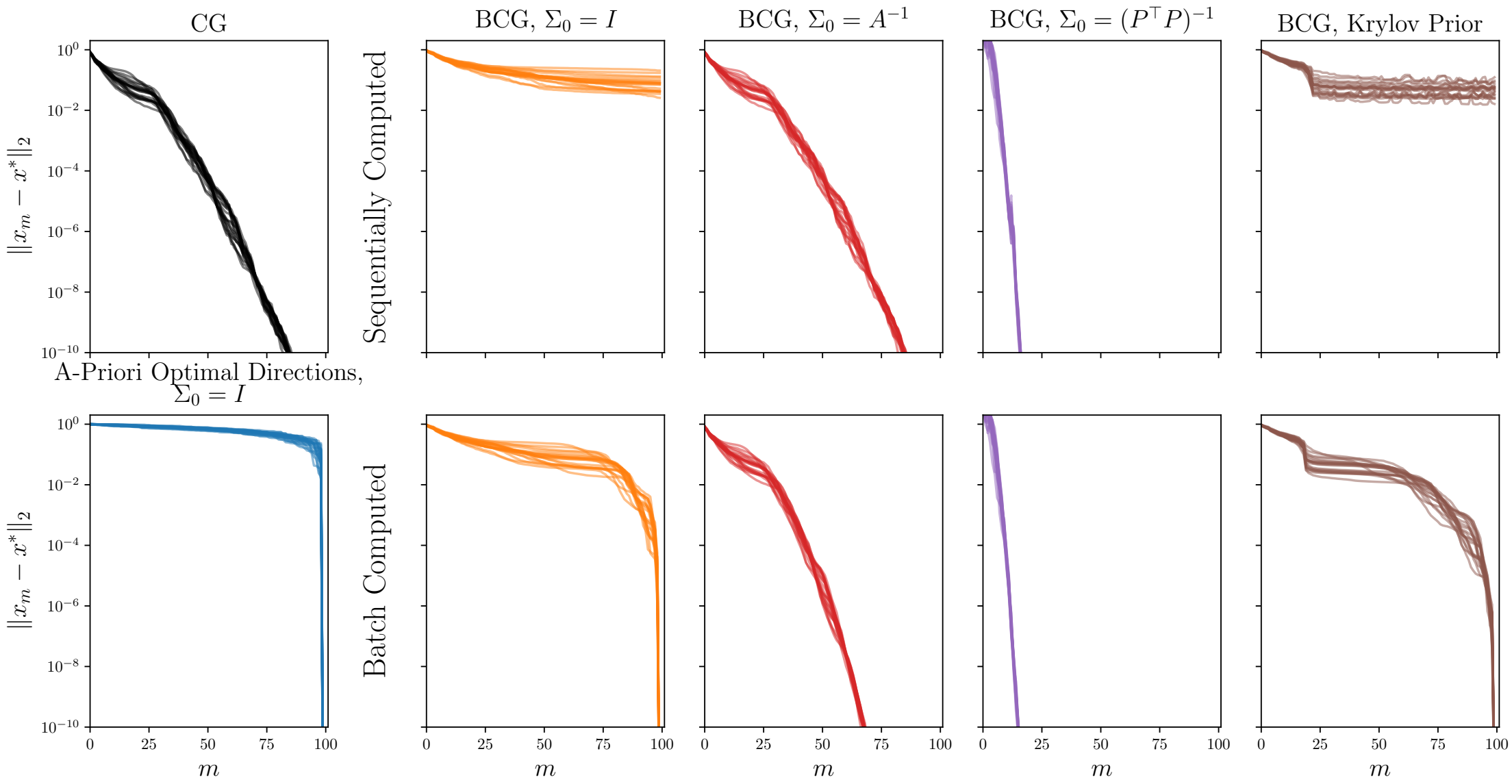

Figure 1: Convergence in mean of BayesCG (BCG). For several independent test problems, $\boldsymbol{x}^* \sim \mu_{\text{ref}}$, the error $\|\boldsymbol{x}_m - \boldsymbol{x}^*\|_2$ was computed. The standard CG method (top left) was compared to variants of BayesCG (right), corresponding to different prior covariances $\Sigma_0$. The search directions used for BayesCG were either computed sequentially (top right) or in batch (bottom right). For comparison, the *a priori* optimal search directions for BayesCG are shown in the bottom left panel.

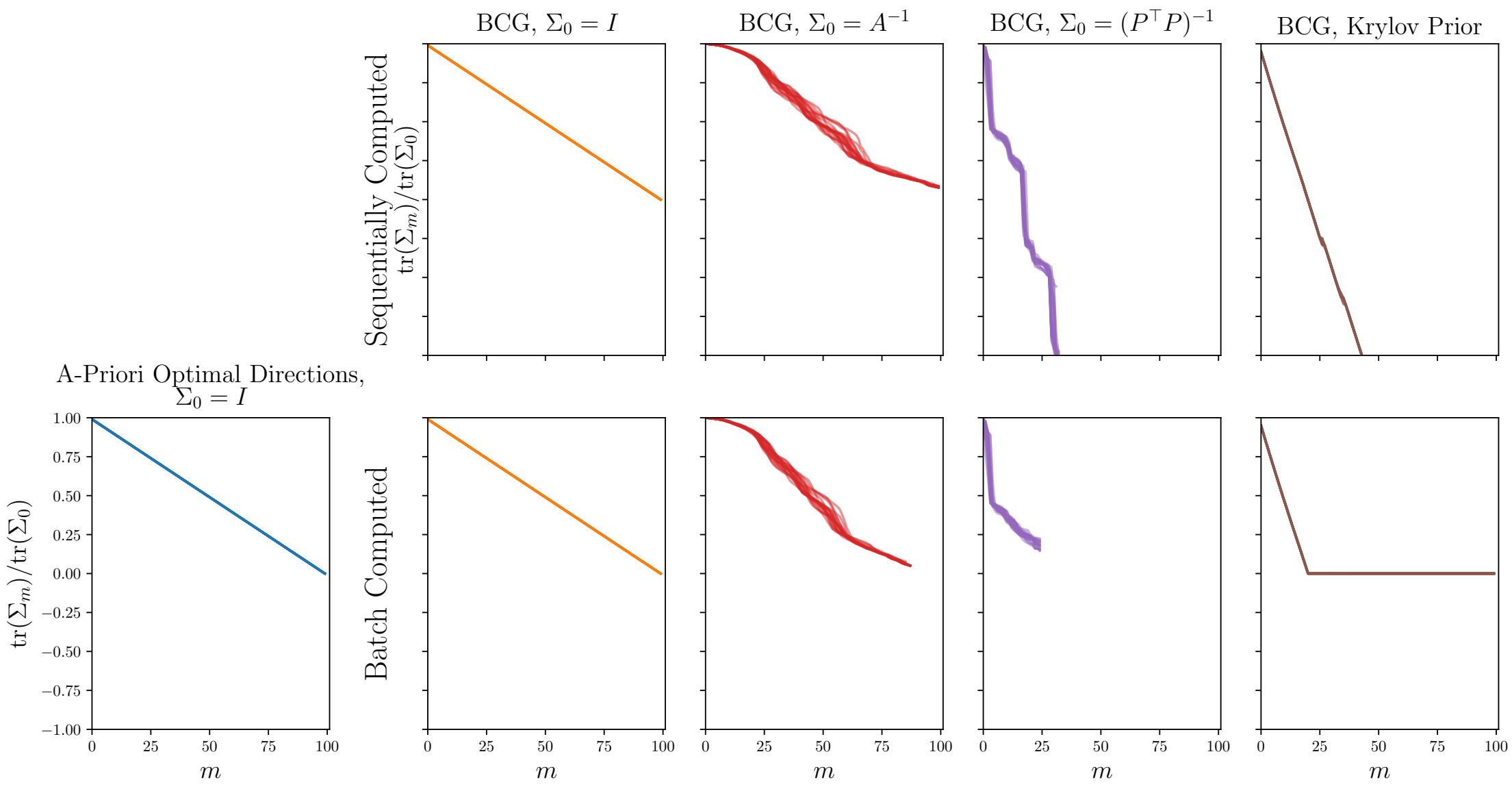

Figure 2: Convergence in posterior covariance of BayesCG (BCG), as measured by $\mathrm{tr}(\Sigma_m)$. The experimental setup was as in Figure 1, here with $\mathrm{tr}(\Sigma_m)/\mathrm{tr}(\Sigma_0)$ plotted.

**Posterior Covariance** In this section the full posterior output from BayesCG is evaluated. In Figure 2, the convergence rate of $\mathrm{tr}(\Sigma_m)$ is plotted for the same set of problems just described to numerically verify the result presented in Proposition 3. Note that while Figure 1 has its $y$-axis on a log-scale, Figure 2 uses a linear scale. It is clear that when the more informative CG or BayesCG search directions are used, the rate of contraction in the posterior mean does not transfer to the posterior covariance. In the remaining columns of the figure, $\mathrm{tr}(\Sigma_m)$ appears to contract at a roughly linear rate, in contrast to the exponential rate observed for $\boldsymbol{x}_m$. This indicates that tightening the bound provided in Proposition 3 is unlikely to be possible. Furthermore, in the last two columns of Figure 2, the impact of numerical non-conjugacy is apparent as the posterior covariance takes on negative values at around $m = 20$.

**Uncertainty Quantification** We now turn to an assessment of the quality of the uncertainty quantification (UQ) being provided. The same experimental setup was used as in the previous sections, however rather than running each variant of BayesCG to $m = d$, instead we ran these until $m = 10$ to ensure that uncertainty quantification (UQ) is needed. To avoid the issue of negative covariances seen in Figure 2, the batch-computed search directions were used throughout.

First, the Gaussian version of BayesCG from Proposition 6 was evaluated. To proceed we used the following argument: When the UQ is well-calibrated, we could consider $\boldsymbol{x}^*$ as plausibly being drawn from the posterior distribution $\mathcal{N}(\boldsymbol{x}_m, \Sigma_m)$. Note that $\Sigma_m$ is of rank $d-m$, but assessing uncertainty in its null space is not of interest as in this space $\boldsymbol{x}^*$ has been determined exactly. Since $\Sigma_m$ is positive semidefinite, it has the singular-value decomposition

$$\Sigma_m = U \begin{bmatrix} D & 0_{d-m,m} \\ 0_{m,d-m} & 0_{m,m} \end{bmatrix} U^\top$$

where $0_{m,n}$ denotes an $m \times n$ matrix of zeroes, $D \in \mathbb{R}^{(d-m)\times(d-m)}$ is diagonal and $U \in \mathbb{R}^{d\times d}$ is an orthogonal matrix. The first $d-m$ columns of $U$, denoted $U_{d-m}$, form a basis of $\mathrm{range}(\Sigma_m)$, the subspace of $\mathbb{R}^d$ in which $\boldsymbol{x}^*$ is still uncertain. Under this hypothesis we can therefore derive a test statistic

$$\begin{aligned} & U_{d-m} D^{-\frac{1}{2}} U_{d-m}^\top (\boldsymbol{x}^* - \boldsymbol{x}_m) \sim \mathcal{N}(\mathbf{0}, I_{d-m}) \\ \implies \quad & Z(\boldsymbol{x}^*) := \|D^{-\frac{1}{2}} U_{d-m}^\top (\boldsymbol{x}^* - \boldsymbol{x}_m)\|_2^2 \sim \chi^2_{d-m} \end{aligned}$$

where here $I_n \in \mathbb{R}^{n\times n}$ is the identity matrix. Note that the pre-factor $U_{d-m}$ is not necessary in the final expression as this norm is unitarily invariant.

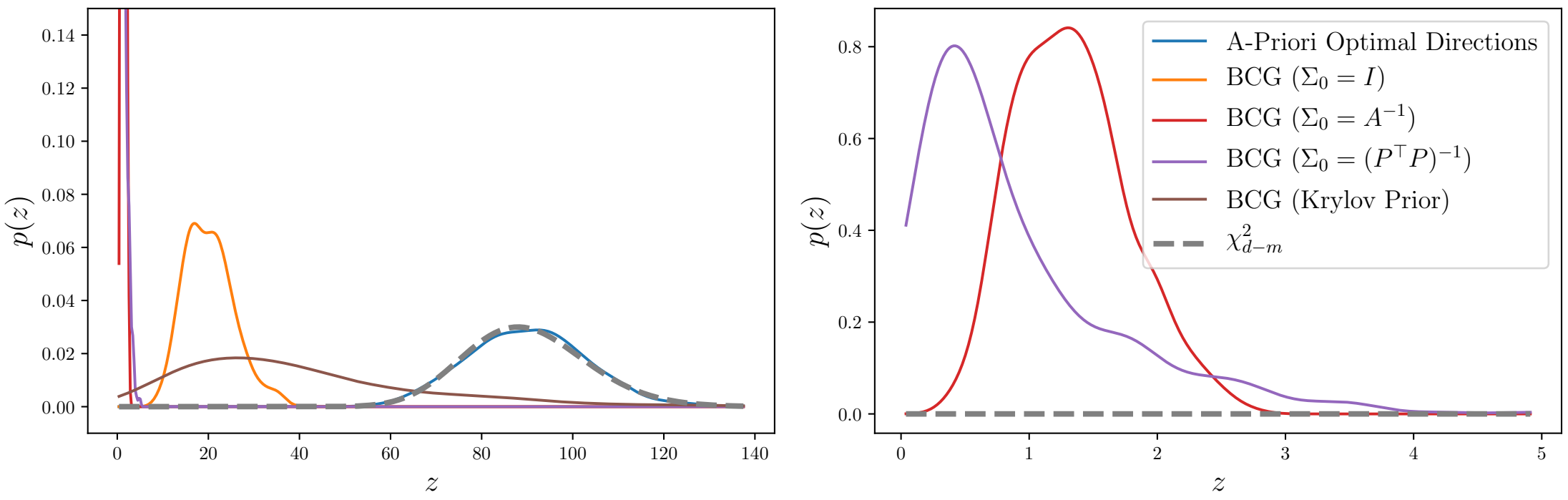

Figure 3: Assessment of the uncertainty quantification provided by the Gaussian BayesCG method, with different choices for search directions and $\Sigma_0$. Plotted are kernel density estimates for the statistic $Z$ based on 500 randomly sampled test problems. These are compared with the theoretical distribution of $Z$ when the posterior distribution is well-calibrated. The right panel zooms in on the estimate for $\Sigma_0 = A^{-1}$ and $\Sigma_0 = (P^\top P)^{-1}$.

Thus to evaluate the UQ we can draw many test problems $\boldsymbol{x}^* \sim \mu_{\text{ref}}$, evaluate the test statistic $Z(\boldsymbol{x}^*)$ and compare the empirical distribution of this statistic to $\chi^2_{d-m}$. If the posterior distribution is well-calibrated we expect that the empirical distribution of the test statistic will resemble $\chi^2_{d-m}$. An overly-conservative posterior will exhibit a "left-shift" in its density, as $\boldsymbol{x}_m$ is closer to $\boldsymbol{x}^*$ than was expected. Likewise, an overly confident posterior will exhibit a "right-shift".

In Figure 3 the empirical distribution of the statistic $Z$ was compared to its theoretical distribution for different prior covariances. The empirical distributions were plotted as kernel density estimates based upon the computed statistic for 500 sampled test problems. Clearly the *a priori* optimal directions provide well-calibrated UQ, while for BayesCG the UQ provided by the posterior was overly-conservative for the prior covariances $\Sigma_0 = I$, $A^{-1}$ and $(P^\top P)^{-1}$. This reflects the fact that the search directions encode knowledge of $\boldsymbol{b}$, but this knowledge is not reflected in the likelihood model used for conditioning, as discussed following Proposition 7. Furthermore, note that the quality of the UQ seems to worsen as the convergence rate for $\boldsymbol{x}_m$ improves, with $\Sigma_0 = (P^\top P)^{-1}$ providing the most conservative UQ.

For the Krylov subspace prior, which encodes intuition for how search directions are selected, better UQ was provided. Though the empirical distribution of $Z$ is not identical to the theoretical distribution, the supports of the two distributions overlap. Thus, while the Krylov subspace prior does not fully remedy the issue caused by the use of $\boldsymbol{b}$ in the search directions, some improvement is seen through the incorporation of knowledge of $\boldsymbol{b}$ into the prior.

Next we assessed the UQ provided by the multivariate $t$ posterior presented in Proposition 11. A similar procedure was followed to the Gaussian case, with a different test statistic. Let $S \sim \mathcal{N}(\mathbf{0}, I)$, $T \sim \mathsf{MVT}_m(\boldsymbol{\mu}, \Sigma)$ and $U \sim \chi^2_m$. Then, it can be shown that

$$\frac{1}{\sqrt{m}} U_{d-m} D^{-\frac{1}{2}} U_{d-m}^\top (T - \boldsymbol{\mu}) \overset{d}{=} \frac{S}{\sqrt{U}} \implies \frac{1}{m} \| D^{-\frac{1}{2}} U_{d-m}^\top (T - \boldsymbol{\mu}) \|_2^2 \overset{d}{=} \frac{\|S\|_2^2}{U}$$

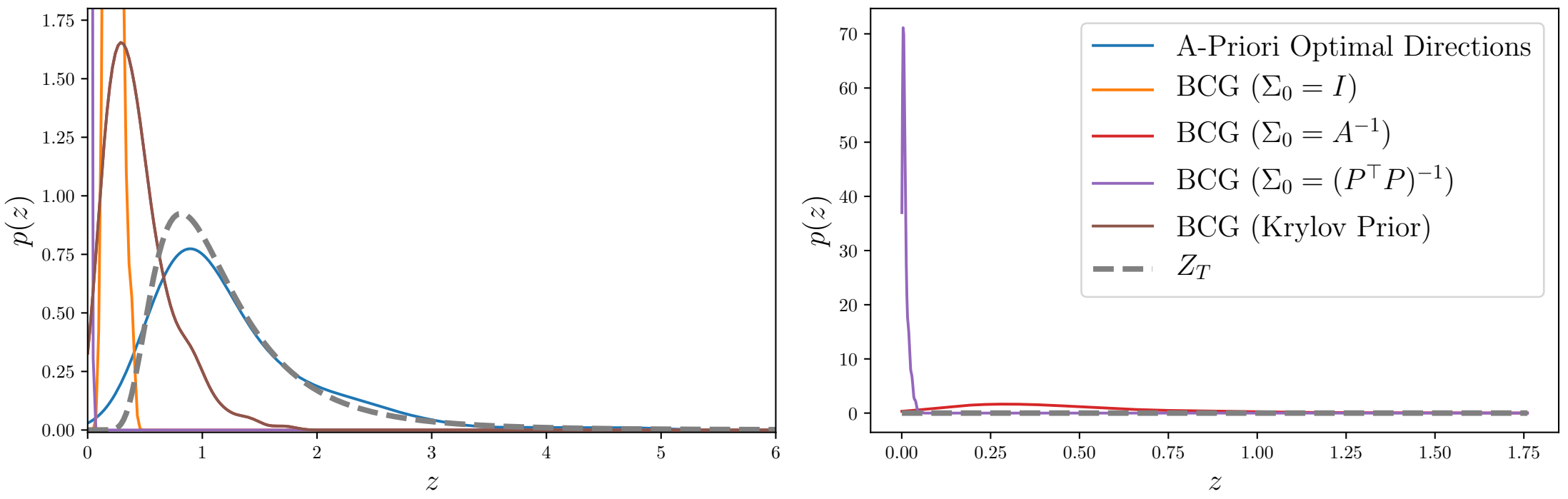

Figure 4: Assessment of the uncertainty quantification provided by the multivariate $t$ BayesCG method, for the same prior covariances and search directions as in Figure 3.

In the present setting, $\boldsymbol{\mu} = \boldsymbol{x}_m$ and $\Sigma = \Sigma_m$. Furthermore $\|S\|_2^2 \sim \chi^2_{d-m}$. Lastly, multiplying both sides by $m/(d-m)$ we have

$$Z(\boldsymbol{x}^*) := \frac{1}{d-m} \| D^{-\frac{1}{2}} U_{d-m}^\top (\boldsymbol{x}_m - \boldsymbol{x}^*) \|_2^2 \overset{d}{=} \frac{\frac{\|S\|_2^2}{(d-m)}}{\frac{U}{m}}.$$

The ratio on the right-hand-side is known to follow an $F(d-m, m)$ distribution. In Figure 4 the empirical distribution of the test statistic $Z(\boldsymbol{x}^*)$ was compared to the $F(d-m, m)$ distribution for each of the posterior distributions considered. Again, the posterior distribution based on the *a priori* optimal search directions was well-calibrated, while the posteriors from BayesCG trade fast convergence in mean with well-calibrated UQ. As before, BayesCG with the Krylov subspace prior appears to provide the best-calibrated UQ of the (practically useful) priors considered.

Note that in both Fig. 3 and Fig. 4, for the choice $\Sigma_0 = (P^\top P)^{-1}$, which has the most rapidly converging mean in Fig. 1, poor UQ properties are observed, making this otherwise appealing choice impractical. To address this we have explored a heuristic procedure for setting $\nu_m$, which aims to match the posterior spread to an appropriate estimate of the error $\|\boldsymbol{x}_m - \boldsymbol{x}^*\|_2$. This procedure is reported in Section S3, along with experimental results based upon it.

**Comparison to Earlier Work** In this section BayesCG is compared to the method proposed in Hennig [2015], which is also briefly recalled here; for full details see the cited work. Rather than performing inference on $\boldsymbol{x}^*$, Hennig [2015] proposed to treat the matrix inverse $A^{-1}$ as the unknown object. Let $\vec{A} \in \mathbb{R}^{d^2}$ denote the vectorisation of $A$,

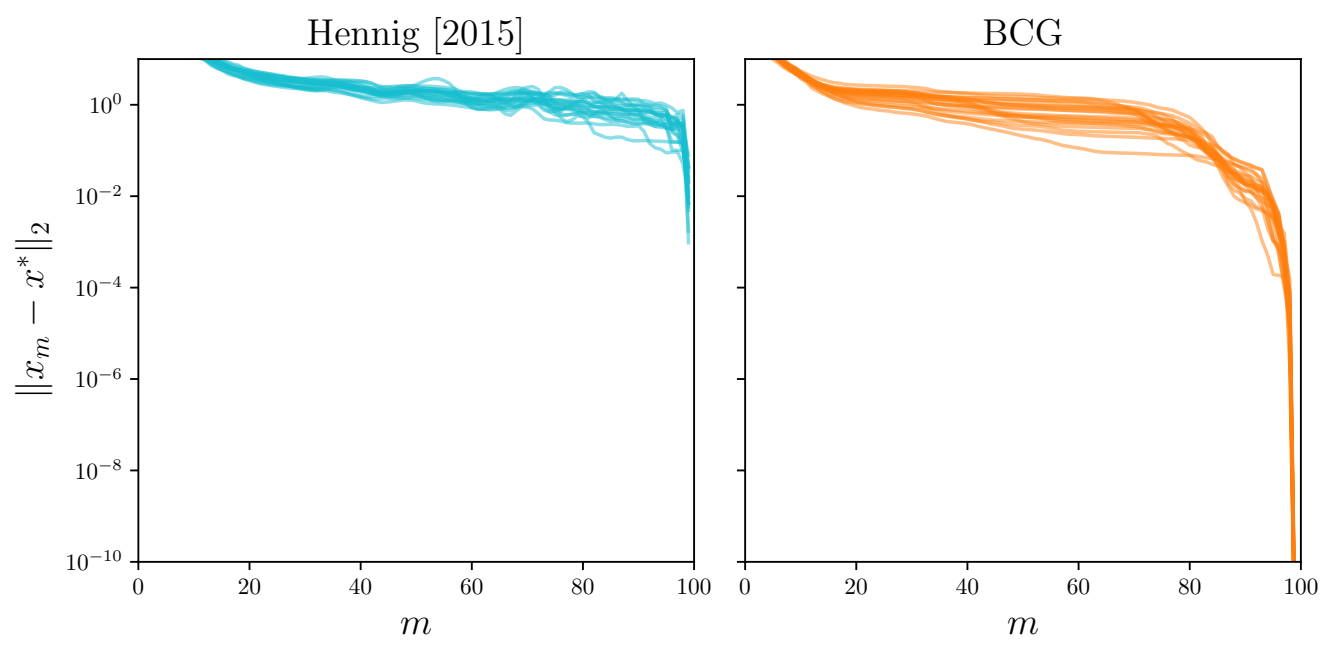

Figure 5: Comparison of the point estimates $\boldsymbol{x}_m$ of Hennig [2015] and the proposed BayesCG method.

defined to be the column vector formed by stacking the transposed rows of $A$ vertically; that is:

$$A = \begin{bmatrix} \boldsymbol{v}_1^\top \\ \vdots \\ \boldsymbol{v}_d^\top \end{bmatrix} \implies \vec{A} = \begin{bmatrix} \boldsymbol{v}_1 \\ \vdots \\ \boldsymbol{v}_d \end{bmatrix}$$

Assume that $A$ is symmetric positive definite and let $H = A^{-1}$. Then a Gaussian prior was placed on $\vec{H}$ with mean $\vec{H_0}$ and covariance $W \circledast W$, the symmetric Kronecker product of a symmetric postive-definite matrix $W \in \mathbb{R}^{d \times d}$ with itself. The observations for the inference were defined by search directions $S_m$ and data $Y_m$, each a matrix in $\mathbb{R}^{d \times m}$ and such that $AS_m = Y_m$. The posterior distribution $H|Y_m$ is then given by $\vec{H} \sim \mathcal{N}(\vec{H_m}, W_m \circledast W_m)$, where:

$$\begin{aligned} H_m &:= H_0 + (S - H_0Y)(Y^\top WY)^{-1}Y^\top W + WY(Y^\top WY)^{-1}(S - H_0Y)^\top \\ &\quad - WY(Y^\top WY)^{-1}[Y^\top(S - H_0Y)](Y^\top WY)^{-1}Y^\top W \\ W_m &:= W - WY(Y^\top WY)^{-1}Y^\top W \end{aligned}$$

where $W_m$ is another symmetric positive definite matrix. To connect the approach of Hennig [2015] to our work, observe that a probability model for $A$ in $A\boldsymbol{x} = \boldsymbol{b}$ induces a probability model for $\boldsymbol{x}$ as the quantities are deterministically coupled:

$$\boldsymbol{x} = H\boldsymbol{b} \sim \mathcal{N}\left(H_m\boldsymbol{b}, \frac{1}{2}(\boldsymbol{b}^\top W_m\boldsymbol{b} \cdot W_m + W_m\boldsymbol{b}\boldsymbol{b}^\top W_m)\right).$$

Thus, projection of the matrix-valued posterior given in Hennig [2015] onto the solution space $\mathbb{R}^d$ is straightforward and does not involve costly computation of the symmetric Kronecker product.

For the experiments that follow, the prior distribution was taken to be $H_0 = W = I$. The system $A$ was a sparse symmetric positive-definite matrix generated randomly as previously described. The true solution $\boldsymbol{x}^*$ was drawn from a multivariate Gaussian distribution $\mathcal{N}(\boldsymbol{0}, I)$, again as described. To ensure a fair comparison, the prior distribution for BayesCG was taken to be the projection of the matrix-valued prior of Hennig [2015] onto the solution space $\mathbb{R}^d$, i.e. $\boldsymbol{x}_0 = \boldsymbol{b}$ and $\Sigma_0 = \boldsymbol{b}^\top \boldsymbol{b} \cdot I + \boldsymbol{b}\boldsymbol{b}^\top$. Note that Hennig [2015] recommend to use an implicit prior to enforce concordance with the CG method. In this experiment the goal was simply to compare the posterior distributions produced by the two approaches, so an arbitrary but equivalent prior was sufficient. The search directions used to form the posterior in Hennig et al. [2015] were taken to be the same as the search directions output by BayesCG. Batch-computed search directions were again used. Thus, the posteriors for the two methods are based upon the same prior assumptions at the level of $\boldsymbol{x}^*$. The information provided to each method is given by the same search directions, but is not equivalent as the projection applied to the search directions is different for each method.

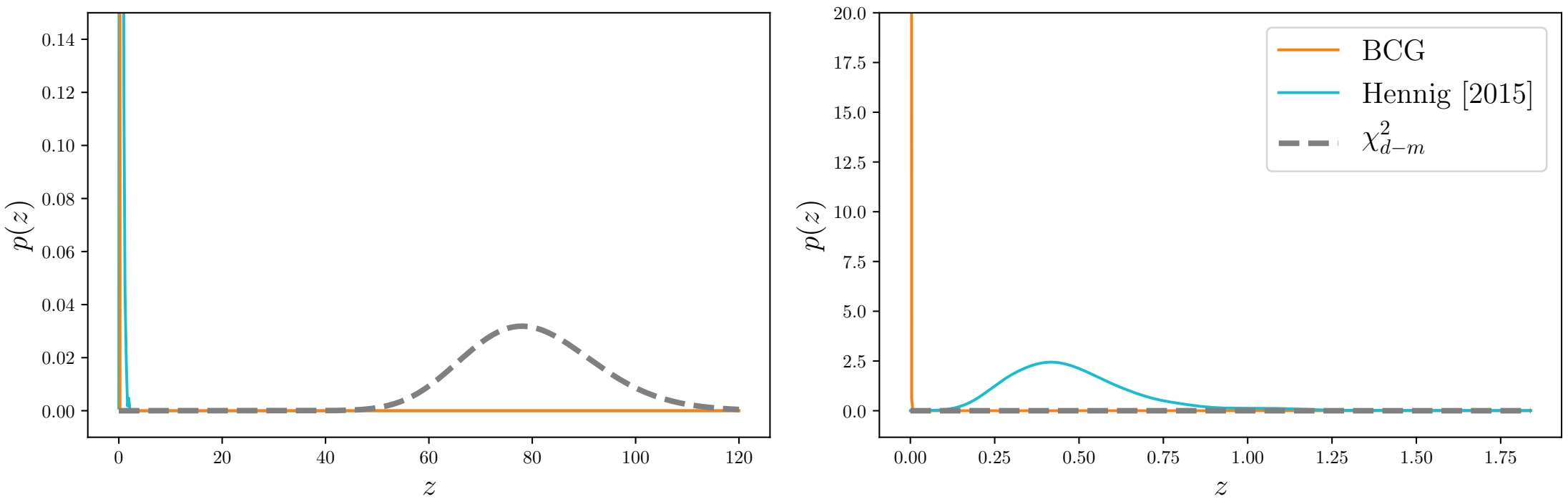

Figure 6: Comparison of the uncertainty quantification provided by BayesCG and the method of Hennig [2015].

In Figure 5 the convergence of $\boldsymbol{x}_m$ from BayesCG was compared to that of the posterior mean in the solution space implied by the matrix-valued method. As might be expected, the convergence rate was approximately the same between the two methods, though the posterior means are not identical. This is due to the aforementioned fact that the information in each example is not equivalent. In Hennig [2015] the information is $S_m = A^{-1} Y_m$, while in BayesCG it is given by $S_m^\top A \boldsymbol{x}^* = S_m^\top \boldsymbol{b}$. Thus, the information in BayesCG is obtained by left-multiplying the information in Hennig [2015] by $\boldsymbol{b}^\top$. Since this projection is not invertible, more information is available to the method of Hennig [2015] than BayesCG, so we should not expect the posteriors to be identical. The instability in the left-hand plot is owing to the non-conjugacy of the directions from BayesCG for the matrix-valued method, which introduces a requirement to solve a linear system. Hennig [2015] propose alternative search directions which eliminate this linear system solve, but this would prevent a fair comparison with BayesCG.

In Figure 6 the UQ provided by the Gaussian version of BayesCG and the method of Hennig [2015] were compared using the approach previously described. Again both approaches provide very conservative UQ compared to the theoretical distribution, with each highly peaked close to zero. The empirical density provided by Hennig [2015] appears to be slightly closer to the theoretical distribution, but both are far from where the theoretical distribution is concentrated so that the difference is inconsequential.

However, while the properties of the two methods for equivalent priors is empirically similar, the range of prior covariance structures introduced and examined in this paper is more difficult to include in the method of Hennig [2015]. In particular, it is less clear how the information exploited in the Krylov subspace prior should be used to inform prior choice in the approach of Hennig [2015], as this knowledge about the solution space rather than $A^{-1}$. Furthermore, while knowledge of $P^{-1}$ provides information about the *location* of $A^{-1}$, it is unclear that this choice of prior mean in Hennig [2015] would result in improved performance as it does in BayesCG. Thus, we believe that the increased flexibility and intuition in prior choice makes BayesCG a more attractive method to a user than the method of Hennig [2015].

## 6.2 Electrical Impedance Tomography

Electrical impedance tomography (EIT) is an imaging technique used to estimate the internal conductivity of an object of interest [Somersalo et al., 1992]. This conductivity is inferred from measurements of voltage induced by applying stimulating currents through electrodes attached to its boundary. EIT was originally proposed for medical applications as a non-invasive diagnostic technique [Holder, 2004], but it has also been applied in other fields, such as engineering [Oates et al., 2017].

The physical relationship between the inducing currents and resulting voltages can be described by a PDE, most commonly the complete electrode model (CEM) [Cheng et al., 1989]. Consider a domain $D \subset \mathbb{R}^n$ representing the object of interest, where typically $n = 2$ or $n = 3$. Denote by $\partial D$ the boundary of $D$, and let $\sigma(\boldsymbol{z})$ denote the conductivity field of interest, where $\boldsymbol{z} \in D$. Denote by $\{e_l\}_{l=1}^{L}$ the $L$ electrodes, where each $e_l \subset \partial D$ and $e_l \cap e_m = \emptyset$ whenever $l \neq m$. Let $v(\boldsymbol{z})$ denote the voltage field, and let $\{I_{i,l}\}_{l=1}^{L}$ denote the set of stimulating currents applied to the electrodes. Let $\{V_{i,l}^{\sigma}\}_{l=1}^{L}$ denote the corresponding voltages, and let $\boldsymbol{n}$ denote the outward-pointing normal vector on $\partial D$. The subscript $i$ here is to distinguish between multiple *stimulation patterns* which are generally applied in sequence and are of relevance to the inversion problem for determining $\sigma(\boldsymbol{z})$ later. Denote by $\{\zeta_l\}_{l=1}^{L}$ the contact impedance of each electrode. The contact impedances are used to model the fact that the contact between the electrode and the boundary of the domain is imperfect. Then the CEM is given by

$$\begin{aligned} -\nabla \cdot (\sigma(\boldsymbol{z}) \nabla v(\boldsymbol{z})) &= 0 && \boldsymbol{z} \in D \\ \int_{e_l} \sigma(\boldsymbol{z}) \frac{\partial v}{\partial \boldsymbol{n}}(\boldsymbol{z}) \mathrm{d}\boldsymbol{z} &= I_{i,l} && l = 1, \dots, L \\ \sigma(\boldsymbol{z}) \frac{\partial v}{\partial \boldsymbol{n}}(\boldsymbol{z}) &= 0 && \boldsymbol{z} \in \partial D \setminus \bigcup_{l=1}^{L} e_l \\ v(\boldsymbol{z}) + \zeta_l \sigma(\boldsymbol{z}) \frac{\partial v}{\partial \boldsymbol{n}}(\boldsymbol{z}) &= V_{i,l}^{\sigma} && \boldsymbol{z} \in e_l,\ l = 1, \dots, L. \end{aligned} \tag{13}$$

A solution of this PDE is the tuple $(v(\boldsymbol{z}), V_{i,1}^{\sigma}, \dots, V_{i,L}^{\sigma})$, consisting of the interior voltage field and the voltage measurements on the electrodes. The numerical solution of this PDE can be reduced to the solution of a linear system of the form in Eq. (1), as will shortly be explained.

Having specified the PDE linking stimulating currents to resulting voltages, it remains to describe the approach for determining $\sigma(\boldsymbol{z})$ from noisy voltage measurements. These physical voltage measurements are denoted by the matrix $V \in \mathbb{R}^{L\times(L-1)}$, where $V_{i,l}$ is the voltage obtained from stimulation pattern $i$ at electrode $l$. The recovery problem can be cast in a Bayesian framework, as formalised in Dunlop and Stuart [2016]. To this end, a prior distribution for the conductivity field is first posited and denoted $\mu_\sigma$. Then, the posterior distribution $\mu_\sigma^V$ is defined through its Radon–Nikodym derivative with respect to the prior as

$$\frac{\mathrm{d}\mu_\sigma^V}{\mathrm{d}\mu_\sigma}(\sigma) \propto \exp(-\Phi(\sigma; V))$$

where $\Phi(\sigma; V)$ is known as a *potential* function and $\exp(-\Phi(\sigma; V))$ is the likelihood. This posterior distribution is for an infinite-dimensional quantity-of-interest and is generically nonparametric, thus sampling techniques such as the preconditioned Crank–Nicolson (pCN) algorithm Cotter et al. [2013] are often employed to access it. Such algorithms require repeated evaluation of $\Phi(\sigma; V)$ and thus the repeated solution of a PDE. Thus, there is interest in ensuring that $\Phi(\sigma; V)$ can be computed at low cost.

**Experimental Setup** The experimental set-up is shown in Figure 8a and is due to Isaacson et al. [2004]. This is described in detail in the supplement. In the absence of specific data on the accuracy of the electrodes, and for convenience, the observational noise was assumed to be Gaussian with standard deviation $\delta = 1$. This implies a potential of the form:

$$\Phi(\sigma; V) = \sum_{i=1}^{L-1}\sum_{l=1}^{L} \frac{(V_{i,l} - V_{i,l}^\sigma)^2}{2\delta^2} = \frac{1}{2\delta^2}(\vec{V} - \vec{V}^\sigma)^\top(\vec{V} - \vec{V}^\sigma)$$

where $V^\sigma$ is the matrix with $(i, l)$-entry $V_{i,l}^\sigma$. The notation $\vec{V} \in \mathbb{R}^{L(L-1)}$ denotes the vectorisation of $V$, as introduced in Section 6.1.

Apart from in pathological cases, there is no analytical solution to the CEM and thus evaluating $\Phi(\sigma; V)$ requires an approximate solution of Eq. (13). Here a finite-element discretisation was used to solve the weak form of Eq. (13), as presented in Dunlop and Stuart [2016] and described in more detail in the supplement. This discretisation results in a sparse system of equations $A\boldsymbol{x}^* = \boldsymbol{b}$, where $A$ is in this context referred to as a *stiffness matrix.* To compute $A$ and $\boldsymbol{b}$, standard piecewise linear basis functions were used, and the computations were perfomed using the `FEniCS` finite-element package. A fine discretisation of the PDE will necessarily yield a high-dimensional linear system to be solved. The idea proposed here is to use BayesCG to approximately solve the linear system, and propagate the solver uncertainty from BayesCG into the the inverse problem associated with recovery of the conductivity field. In essence, this provides justification for small values of $m$ to be used in the linear solver and yet ensure that the inferences for $\sigma$ remain valid.

The Gaussian version of BayesCG was used throughout, as described in Proposition 6. Thus, assume that the output from BayesCG is $\boldsymbol{x} \sim \mathcal{N}(\boldsymbol{x}_m, \Sigma_m)$. The finite element approximation to the voltages $V_{i,l}^{\sigma}$ is linearly related to the solution $\boldsymbol{x}^*$ of the linear system, so that BayesCG implies a probability model for the voltages of the form $\vec{V}^{\sigma} \sim \mathcal{N}(\vec{V}_m^{\sigma}, \Sigma_m^{\sigma})$ for some $\vec{V}_m^{\sigma}$ and $\Sigma_m^{\sigma}$; for brevity we leave these expressions implicit. The approach proposed is to derive a new potential $\hat{\Phi}$, obtained by marginalising the posterior distribution output from BayesCG in the likelihood. It is straightforward to show that, for the Gaussian likelihood, this marginalisation results in the new potential

$$\hat{\Phi}(\sigma; V) = \frac{1}{2}(\vec{V} - \vec{V}_m^{\sigma})^{\top}(\Sigma_m^{\sigma} + \delta^2 I)^{-1}(\vec{V} - \vec{V}_m^{\sigma}).$$

Thus, the new likelihood $\exp(-\hat{\Phi}(\sigma; V))$ is still Gaussian, but with a covariance inflated by $\Sigma_m^{\sigma}$ to account for the level of inaccuracy in the BayesCG solver. It will be shown that replacing $\Phi$ with $\hat{\Phi}$ leads in turn to a posterior distribution $\hat{\mu}_{\sigma}^{V}$ for the conductivity field which is appropriately widened to account for the additional uncertainty modelled in BayesCG.

Throughout this section the prior distribution over the conductivity field was taken to be a centered log-Gaussian distribution, $\log(\sigma) \sim \mathcal{GP}(0, k)$, with a Matérn 5/2 covariance as given by:

$$k(\boldsymbol{z}, \boldsymbol{z}') = a\left(1 + \frac{\sqrt{5}\|\boldsymbol{z} - \boldsymbol{z}'\|_2}{\ell} + \frac{5\|\boldsymbol{z} - \boldsymbol{z}'\|_2^2}{3\ell^2}\right)\exp\left(-\frac{\sqrt{5}\|\boldsymbol{z} - \boldsymbol{z}'\|_2}{\ell}\right).$$

The length-scale parameter $\ell$ was set to $\ell = 1.0$, while the amplitude $a$ was set to $a = 9.0$ to ensure that where the posterior distribution is concentrated has significant probability mass under the prior.

**Forward Problem** Initially the solution to the *forward problem*, i.e. solving the PDE, was considered for a particular stimulation pattern. In Figure 7 the convergence of the point estimates provided by BayesCG and CG were compared, both with the isotropic prior covariance $\Sigma_0 = I$ and with the preconditioner covariance $\Sigma_0 = (P^{\top}P)^{-1}$. For discretising the PDE, we used mesh resolutions $N_d = 64, 128$ and 256, where a larger $N_d$ provides a more accurate discretisation. The precise method of generating the meshes and the meaning of $N_d$ is described in the supplement. For the conductivity field we examined both samples from $\mu_{\sigma}$ and the mean of $\mu_{\sigma}^{V}$, denoted $\hat{\sigma}$ and obtained using MCMC with a fine discretisation of the PDE and an accurate linear solver. As in the previous section, the preconditioner $P$ was given by an incomplete Cholesky factorisation of $A$.

The results in Figure 7 show that when $\Sigma_0 = I$ convergence is slow, but that this is improved when the preconditioner prior is used. This is the same as observed in the previous section, but since this problem is now obtained from an applied example rather than being given by an arbitrary sampled system, it is useful to know that the same observations transfer. The convergence of the estimator from the perspective the conductivity field $\sigma$ is displayed in the supplement.

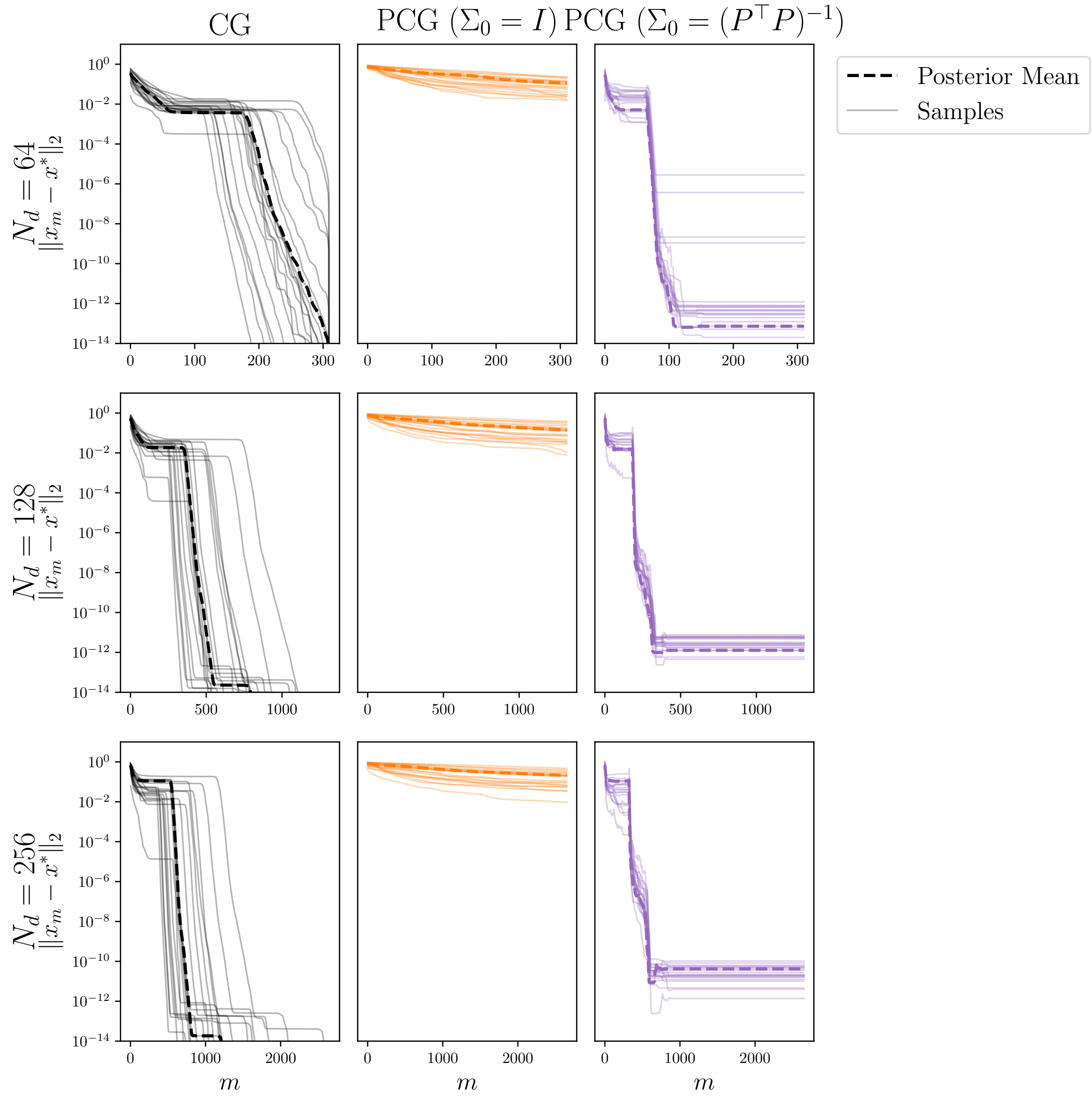

Figure 7: Convergence of the posterior for a linear system arising from a discretisation of the PDE in Eq. (13), for a number of different conductivity fields and discretisation resolutions $N_d$. The solid lines represent the convergence of the BayesCG posterior mean for conductivity fields sampled from the prior $\mu^\sigma$. The dashed lines are for the conductivity field obtained as the the mean of the posterior $\mu_\sigma^V$.

**Inverse Problem** In this section, the solution to the inverse problem when using the BayesCG potential $\hat{\Phi}$ is compared to the posterior obtained from the exact potential $\Phi$. In the latter case CG was used to solve the system to convergence to provide a brute-force benchmark. For BayesCG, the prior was centered, $\boldsymbol{x}_0 = \mathbf{0}$, and the preconditioner prior covariance, $\Sigma_0 = (P^\top P)^{-1}$, was used. BayesCG was run to $m = 80$ iterations,

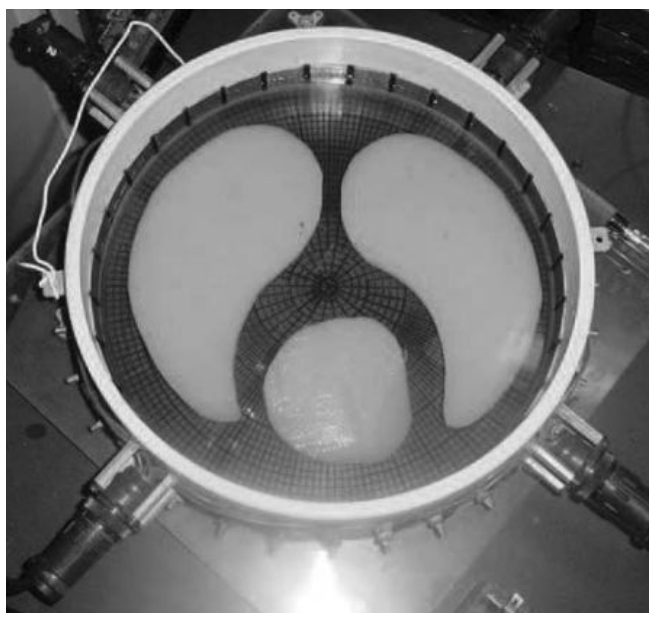

(a) Set-up for the experiment described in Section 6.2.

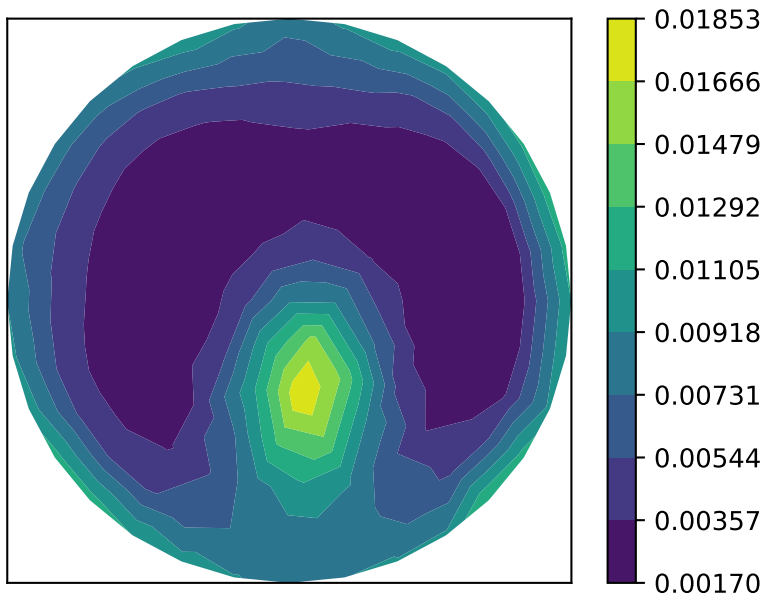

(b) Exact posterior mean for $\log \sigma$

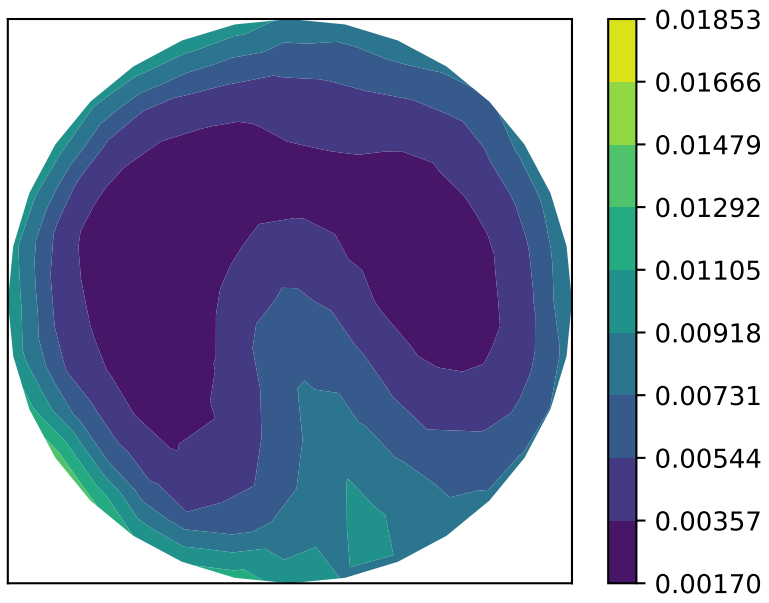

(c) BCG-based posterior mean for $\log \sigma$

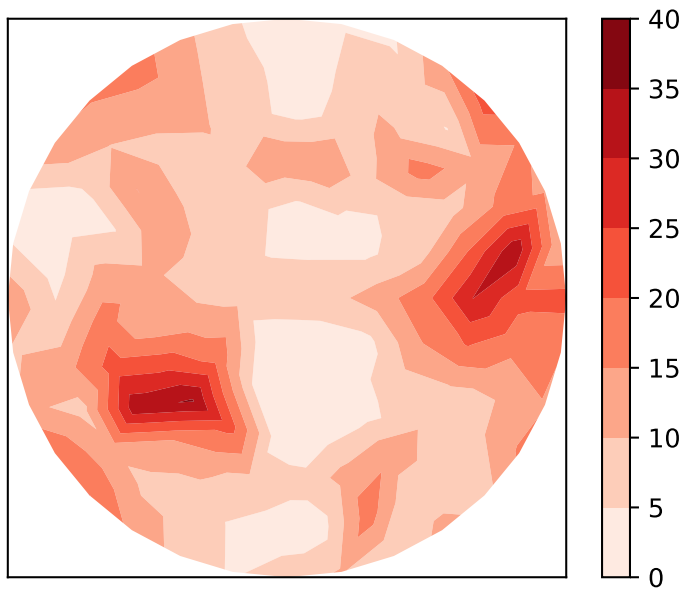

(d) Ratio of point-wise posterior standard deviation, for BayesCG-based compared to exact.

Figure 8: Comparison of the posterior distribution over the conductivity field, when using BayesCG to solve the linear system arising from the forward problem compared to using standard CG.

for the mesh with $N_d = 64$. This mesh results in a linear system with $d = 311$, so 80 iterations represents a relatively small amount of computational effort.

In Figure 8 the posterior distribution over the conductivity field is displayed. In Figures 8b and 8c, respectively, the exact posterior mean and the posterior mean from BayesCG are plotted. Note that, as indicated in the previous section, many of the features of the conductivity field have been recovered even though a relatively small number of iterations have been performed. In Figure 8d the ratio of the pointwise posterior standard deviation from BayesCG to that in CG is plotted. Clearly, throughout the entire spatial domain, the posterior distribution has a larger standard deviation, showing that the posterior uncertainty from BayesCG has successfully been transferred to the posterior over the conductivity field. This results in a posterior distribution which is wider to account for the fact that an imperfect solver was used to solve the forward problem. Overall, the integrated standard deviation over the domain is 0.0365 for BayesCG, while for the exact posterior it is 0.0046.

This example illustrates how BayesCG could be used to relax the computational effort required in EIT in such a way that the posterior is widened to account for the imperfect solution to the forward problem. This setting, as well as other applications of this method, should be explored in more detail in future work.

# 7 Conclusion and Discussion

In this paper we have introduced and theoretically analysed the Bayesian conjugate gradient method, a Bayesian probabilistic numerical method for the solution of linear systems of equations. Given the ubiquity of linear systems in numerical computation, the question of how to approximate their solution is fundamental. Contrary to CG and other classical iterative methods, BayesCG outputs a probability distribution, providing a principled quantification of uncertainty about the solution after exploring an $m$-dimensional subspace of $\mathbb{R}^d$. Through the numerical example in Section 6.2 we have shown how this output could be used to make meaningful inferences in applied problems, with reduced computational cost in terms of iterations performed. This could be applied to a broad range of problems in which solution of large linear systems is a bottleneck, examples of which have been given Section 1.1

**Prior Choice** Prior choice was discussed in detail. An important question that arises here is to what extent the form of the prior can be relaxed. Indeed, in many applied settings information is known about $\boldsymbol{x}^*$ which cannot be encoded into a Gaussian prior. A common example of this arises in the solution of PDEs, when it is often the case that the true solution to the PDE is sign-constrained. However, in encoding additional prior information it is likely that the conjugacy properties exploited to construct a closed-form posterior will be be lost. Then, interrogating such posteriors would require sampling techniques such as the numerical disintegration procedure of Cockayne et al. [2017], which would incur a dramatically higher cost. Research to determine what prior knowledge can be encoded (either exactly or approximately) without sacrificing numerical performance will be an important future research direction.

It was shown how a numerical analyst's intuition that the conjugate gradient method "tends to work well" can be encoded into a Krylov-based prior. This went some way towards compensating for the fact that the search directions in BayesCG are constructed in a data-driven manner which is not explicitly acknowledged in the likelihood. Alternative heuristic procedures for calibrating the UQ were explored in the supplement, Section S3. An important problem for future research will be to provide practical and theoretically justified methods for ensuring the posterior UQ is well-calibrated.

**Computational Cost and Convergence** The computational cost of BayesCG is only a constant factor higher than that of CG. However, the convergence rates reported in Section 3 can be slower than those of CG. To achieve comparable convergence rates, the prior covariance $\Sigma_0$ must be chosen to counteract the fact that the rate is based on $\kappa(\Sigma_0 A^\top A)$ rather than $\kappa(A)$, and this can itself incur a substantial computational cost. Future work will focus on reducing the cost associated with BayesCG.

**Acknowledgements** Chris J. Oates and Mark Girolami were supported by the Lloyd's Register Foundation programme on Data-Centric Engineering. Ilse C.F. Ipsen was supported in part by NSF grant DMS-1760374. Mark Girolami was supported by EPSRC grants [EP/R034710/1, EP/R018413/1, EP/R004889/1, EP/P020720/1], an EPSRC Established Career Fellowship [EP/J016934/3] and a Royal Academy of Engineering Research Chair.

This material was based upon work partially supported by the National Science Foundation under Grant DMS-1127914 to the Statistical and Applied Mathematical Sciences Institute. Any opinions, findings, and conclusions or recommendations expressed in this material are those of the author(s) and do not necessarily reflect the views of the National Science Foundation.

The authors are grateful to Shewchuk [1994], which provides an accessible introduction to CG. Techniques described therein formed the basis of some of the proofs in this paper. They would also like to thank Philipp Hennig and Simon Bartels for their help in the implementation of their algorithm and Tim Sullivan for his feedback on the manuscript.

# Supplementary Material for “A Bayesian Conjugate-Gradient Method”

Jon Cockayne[*], Chris J. Oates[†], Ilse C.F. Ipsen[‡] and Mark Girolami[§]

December 17, 2018

## S1 Proof of Theoretical Results

*Proof of Proposition 1.* Note that the joint distribution of $\boldsymbol{x}$ and $\boldsymbol{y}_m$ is given by

$$\begin{bmatrix} \boldsymbol{x} \\ \boldsymbol{y}_m \end{bmatrix} \sim \mathcal{N}\left( \begin{bmatrix} \boldsymbol{x}_0 \\ S_m^\top A \boldsymbol{x}_0 \end{bmatrix}, \begin{bmatrix} \Sigma_0 & \Sigma_0 A^\top S_m \\ S_m^\top A \Sigma_0 & S_m^\top A \Sigma_0 A^\top S_m \end{bmatrix} \right)$$

from which the stated conditional distribution is deduced. □

*Proof of Proposition 2.* Consider an arbitrary vector $\boldsymbol{\ell} \in \mathbb{R}^d$. Now

$$\begin{aligned} \boldsymbol{\ell}^\top \boldsymbol{x}_m - \boldsymbol{\ell}^\top \boldsymbol{x}^* &= \boldsymbol{\ell}^\top (\boldsymbol{x}_0 - \boldsymbol{x}^*) + \boldsymbol{\ell}^\top \Sigma_0 A^\top S_m \Lambda_m^{-1} S_m^\top A (\boldsymbol{x}^* - \boldsymbol{x}_0) && \text{(from Eq. 5)} \\ &= \boldsymbol{\ell}^\top (\Sigma_0 - \Sigma_0 A^\top S_m \Lambda_m^{-1} S_m^\top A \Sigma_0) \Sigma_0^{-1} (\boldsymbol{x}_0 - \boldsymbol{x}^*) \\ &= \langle \Sigma_m \boldsymbol{\ell}, \boldsymbol{x}_0 - \boldsymbol{x}^* \rangle_{\Sigma_0^{-1}} && \text{(from Eq. 6)} \end{aligned}$$

and so:

$$\begin{aligned} |\boldsymbol{\ell}^\top \boldsymbol{x}_m - \boldsymbol{\ell}^\top \boldsymbol{x}^*| &= \left| \langle \Sigma_m \boldsymbol{\ell}, \boldsymbol{x}_0 - \boldsymbol{x}^* \rangle_{\Sigma_0^{-1}} \right| \\ &\leq \|\boldsymbol{x}_0 - \boldsymbol{x}^*\|_{\Sigma_0^{-1}} \underbrace{\|\Sigma_m \boldsymbol{\ell}\|_{\Sigma_0^{-1}}}_{(*)} . \end{aligned} \tag{S1}$$

[*]University of Warwick (j.cockayne@warwick.ac.uk).
[†]Newcastle University and the Alan Turing Institute (chris.oates@ncl.ac.uk).
[‡]North Carolina State University (ipsen@ncsu.edu)
[§]Imperial College London and the Alan Turing Institute (m.girolami@imperial.ac.uk).

where the last line follow from Cauchy–Schwarz. Now, by expanding the term $(*)$ and simplifying, we see that

$$\begin{aligned}
\|\Sigma_m \boldsymbol{\ell}\|^2_{\Sigma_0^{-1}} &= \boldsymbol{\ell}^\top (\Sigma_0 - \Sigma_0 A^\top S_m \Lambda_m^{-1} S_m^\top A \Sigma_0)^\top \Sigma_0^{-1} (\Sigma_0 - \Sigma_0 A^\top S_m \Lambda_m^{-1} S_m^\top A \Sigma_0) \boldsymbol{\ell} \\
&= \boldsymbol{\ell}^\top \big(\Sigma_0 - 2\Sigma_0 A^\top S_m \Lambda_m^{-1} S_m^\top A \Sigma_0 \\
&\qquad + \Sigma_0 A^\top S_m \Lambda_m^{-1} \underbrace{S_m^\top A \Sigma_0 A^\top S_m}_{=\Lambda_m} \Lambda_m^{-1} S_m^\top A \Sigma_0 \big) \boldsymbol{\ell} \\
&= \boldsymbol{\ell}^\top (\Sigma_0 - \Sigma_0 A^\top S_m \Lambda_m^{-1} S_m^\top A \Sigma_0) \boldsymbol{\ell} \qquad \text{(S2)} \\
&= \boldsymbol{\ell}^\top \Sigma_m \boldsymbol{\ell}
\end{aligned}$$

which follows from Eq. 6

Finally let $\boldsymbol{e}_i$ denote the vector whose $j^{\text{th}}$ entry is $\delta_{ij}$ and note that

$$\begin{aligned}
\|\boldsymbol{x}_m - \boldsymbol{x}^*\|_{\Sigma_0^{-1}} &= \|\Sigma_0^{-\frac{1}{2}} (\boldsymbol{x}_m - \boldsymbol{x}^*)\|_2 \\
&= \left( \sum_{i=1}^d \left| \boldsymbol{e}_i^\top \Sigma_0^{-\frac{1}{2}} \boldsymbol{x}_m - \boldsymbol{e}_i^\top \Sigma_0^{-\frac{1}{2}} \boldsymbol{x}^* \right|^2 \right)^{\frac{1}{2}} \\
&\le \|\boldsymbol{x}_0 - \boldsymbol{x}^*\|_{\Sigma_0^{-1}} \left( \sum_{i=1}^d \boldsymbol{e}_i^\top \Sigma_0^{-\frac{1}{2}} \Sigma_m \Sigma_0^{-\frac{1}{2}} \boldsymbol{e}_i \right)^{\frac{1}{2}} \qquad \text{(from Eq. S1, S2)} \\
&= \|\boldsymbol{x}_0 - \boldsymbol{x}^*\|_{\Sigma_0^{-1}} \sqrt{\operatorname{tr}\left( \Sigma_0^{-\frac{1}{2}} \Sigma_m \Sigma_0^{-\frac{1}{2}} \right)} \\
&= \|\boldsymbol{x}_0 - \boldsymbol{x}^*\|_{\Sigma_0^{-1}} \sqrt{\operatorname{tr}(\Sigma_m \Sigma_0^{-1})}
\end{aligned}$$

where the last line uses the fact that the trace is invariant under cyclic permutation of the argument. □

*Proof of Proposition 3.* Note that

$$\begin{aligned}
\operatorname{tr}(\Sigma_m \Sigma_0^{-1}) &= \operatorname{tr}(I - \Sigma_0 A^\top S_m \Lambda_m^{-1} S_m^\top A) \\
&= \operatorname{tr}(I) - \operatorname{tr}(\Sigma_0 A^\top S_m \Lambda_m^{-1} S_m^\top A) \\
&= \operatorname{tr}(I) - \operatorname{tr}(\underbrace{S_m^\top A \Sigma_0 A^\top S_m}_{=\Lambda_m} \Lambda_m^{-1}) \\
&= d - m
\end{aligned}$$

where the third line uses the fact that the trace is invariant under cyclic permutation of the argument. □

*Proof of Proposition 4.* First, note that

$$\Lambda_m = (S_m^{\text{CG}})^\top A \Sigma_0 A^\top S_m^{\text{CG}} = (S_m^{\text{CG}})^\top A S_m^{\text{CG}} = I$$

since the columns of $S_m^{\mathrm{CG}}$ are $A$-orthonormal. Then, from Proposition 1 we have

$$\begin{aligned}\boldsymbol{x}_m &= \boldsymbol{x}_0 + \Sigma_0 A^\top S_m^{\mathrm{CG}} \Lambda_m^{-1} (S_m^{\mathrm{CG}})^\top \boldsymbol{r}_0 \\ &= S_m^{\mathrm{CG}} (S_m^{\mathrm{CG}})^\top \boldsymbol{r}_0 \\ &\equiv \boldsymbol{x}_m^{\mathrm{CG}}\end{aligned}$$

as required.

□

*Proof of Proposition 5.* We first introduce the concept of an average-case optimal algorithm and average-case optimal information. The *information space* $B$ and the *solution space* $X$ are, informally, the spaces in which the right-hand-side and the solution of the system live, respectively. We wish to computationally approximate an intractable *solution operator* $\mathcal{A}(b)$, based upon a finite amount of information provided by the *information operator* $S_m : B \to \mathbb{R}^m$. This is accomplished by an *algorithm* $\psi(S_m(b))$, which we hope approximates $\mathcal{A}(b)$ well in a way which will now be made formal.

For a reference measure $\nu$ on $B$, denote the *average-case error* of an algorithm $\psi$ with information $S_m$ as

$$e_M^{\mathrm{avg}}(S_m, \psi) := \left[ \int_{\mathbb{R}^d} \|\mathcal{A}(b) - \psi(S_m(b))\|_M^2 \, \mu(\mathrm{d}\boldsymbol{x}) \right]^{\frac{1}{2}}.$$

An algorithm $\psi^*$ which minimises $e^{\mathrm{avg}}(\cdot, \psi)$ for arbitrary $S_m$ is said to be *average-case-optimal.* An $S_m^*$ which minimises $e^{\mathrm{avg}}(S_m, \psi^*)$ is said to be *average-case optimal information.*

By Theorem 3.3 of Cockayne et al. [2017], in the present setting optimal information for the average risk in Eq. (10) is identical to average-case optimal information. This is by virtue of the fact that, for any symmetric positive-definite $M$, $(\mathbb{R}^d, \langle \cdot, \cdot \rangle_M)$ forms an inner-product space.

Now recall two relevant theorems from Novak and Woźniakowski [2008]. For measurable spaces $(B, \mathcal{F}_B)$ and $(X, \mathcal{F}_B)$, an operator $\mathcal{A} : B \to X$ and a measure $\mu$ on $B$, let $\mathcal{A}_\# \mu$ denote the *pushforward* of $\mu$ through $\mathcal{A}$, a measure on $X$ defined as

$$[\mathcal{A}_\# \mu](C) = \mu(\mathcal{A}^{-1}(C))$$

for each $C \in \mathcal{F}_X$.

**Theorem S1** (Theorem 4.28 of Novak and Woźniakowski [2008])**.** *Let $B$ be a separable real Banach space equipped with a zero-mean Gaussian measure $\nu$ with covariance operator $C_\nu$. Let the solution operator $\mathcal{A} : B \to X$ be a bounded linear operator into a separable real Hilbert space $X$ with inner product $\langle \cdot, \cdot \rangle_X$. Let $\eta = \mathcal{A}_\# \nu$ be a Gaussian measure on solution elements. Consider linear information $S_m = [s_1, \ldots, s_m]$ where $s_i : B \to \mathbb{R}$ and $s_i(C_\nu s_j) = \delta_{ij}$, and consider information $y_i = s_i(b)$. Then the algorithm*

$$\psi(b) = \sum_{i=1}^m y_i \mathcal{A}(C_\nu s_i)$$

*is average-case optimal.*

Denote by $C_\eta$ the covariance operator of $\eta$, and let $\{(\gamma_i^*, \phi_i^*) : i \in I\}$ for $I \subseteq \mathbb{N}$ denote its eigensystem, ordered so that $\gamma_1^* \geq \gamma_2^* \geq \dots$. Note that if $X$ is finite-dimensional with dimension $d$ then $I = \{1, \dots, d\}$, while otherwise $I = \mathbb{N}$.

**Theorem S2** (Theorem 4.30 of Novak and Woźniakowski [2008])**.** *Under the assumptions of Theorem S1, for $b \in B$ the optimal information $S_m^*$ is given by*

$$S_m^*(b) = [L_1^*(b), \dots, L_m^*(b)]$$

*where*

$$L_i^*(b) := \frac{\langle \mathcal{A}(b), \phi_i^* \rangle_X}{(\gamma_i^*)^{\frac{1}{2}}}.$$

We will first establish that the posterior mean from Proposition 1 represents an average-case optimal algorithm, by applying Theorem S1. In the notation of that theorem, $B = X = \mathbb{R}^d$, which satisfies the required assumptions as $\mathbb{R}^d$ is separable. The measure $\nu$ is given by $\nu = A_\# \mu \sim \mathcal{N}(\mathbf{0}, A\Sigma_0 A^\top)$, so that $C_\nu = A\Sigma_0 A^\top$. Furthermore the information operator $S_m$ is simply a matrix in $\mathbb{R}^{d \times m}$, which is subject to the restriction from Theorem S1 that $\Lambda_m = S_m^\top A \Sigma_0 A^\top S_m = I$. Note that this is markedly similar to the conjugacy requirement in Section 2.2.

Now we seek the optimal algorithm $\psi(\boldsymbol{b})$ which minimises

$$\begin{aligned}\int_{\mathbb{R}^d} \|A^{-1}\boldsymbol{b} - \psi(S_m^\top \boldsymbol{b})\|_M^2 \, \nu(\mathrm{d}\boldsymbol{b}) &= \int_{\mathbb{R}^d} \|M^{\frac{1}{2}} A^{-1}\boldsymbol{b} - M^{\frac{1}{2}}\psi(S_m^\top \boldsymbol{b})\|_2^2 \, \nu(\mathrm{d}\boldsymbol{b}) \\ &= \int_{\mathbb{R}^d} \|M^{\frac{1}{2}} A^{-1}\boldsymbol{b} - \bar{\psi}(S_m^\top \boldsymbol{b})\|_2^2 \, \nu(\mathrm{d}\boldsymbol{b}) \qquad \text{(S3)}\end{aligned}$$

where $\bar{\psi} = M^{\frac{1}{2}}\psi$. Eq. (S3) is of the form required by Theorem S1, with the solution operator $\mathcal{A} = M^{\frac{1}{2}} A^{-1}$, which is a bounded linear operator as required. For any $S_m$ conjugate to $A\Sigma_0 A^\top$, the optimal algorithm is therefore given by

$$\begin{aligned}\bar{\psi}(\boldsymbol{b}) &= \sum_{i=1}^m (\boldsymbol{s}_i^\top \boldsymbol{b}) M^{\frac{1}{2}} A^{-1} A \Sigma_0 A^\top \boldsymbol{s}_i \\ &= M^{\frac{1}{2}} \Sigma_0 A^\top S_m S_m^\top \boldsymbol{b} \\ \implies \psi(\boldsymbol{b}) &= \Sigma_0 A^\top S_m S_m^\top \boldsymbol{b}.\end{aligned}$$

In this conjugate setting with $\boldsymbol{x}_0 = \mathbf{0}$, this is identical to the expression for $\boldsymbol{x}_m$ in Proposition 1.

Theorem S2 can now be applied to determine the optimal information $S_m^*$. Note that since $A$ is a bijection, $\eta = [M^{\frac{1}{2}} A^{-1}]_\# [A_\# \mu] = M^{\frac{1}{2}}_\# \mu$, so the required eigensystem is that of $M^{\frac{1}{2}} \Sigma_0 M^{\frac{\top}{2}}$. As before, denote this (ordered) eigensystem by $\{(\gamma_i^*, \boldsymbol{\phi}_i^*)\}_{i=1}^d$, with the eigenvectors normalised so that $(\boldsymbol{\phi}_i^*)^\top \boldsymbol{\phi}_i^* = 1$. It then holds from Theorem S2 that the optimal search directions are given by

$$\boldsymbol{s}_i^* = (\gamma_i^*)^{-\frac{1}{2}} A^{-\top} M^{\frac{\top}{2}} \boldsymbol{\phi}_i^*.$$

Lastly, noting that the scaling by $(\gamma_i^*)^{-\frac{1}{2}}$ does not affect the output yields the result that the optimal information is given by

$$S_m = A^{-\top} M^{\frac{\top}{2}} \Phi_m.$$

□

*Proof of Proposition 6.* First, note that $\Lambda_m = I$ as the search directions $\{\boldsymbol{s}_i\}$, $i = 1, \ldots, m$ are $Q$-orthonormal, where $Q = A\Sigma_0 A^\top$. Then, from Eq. 5:

$$\begin{aligned}
\boldsymbol{x}_m &= \boldsymbol{x}_0 + \Sigma_0 A^\top S_m S_m^\top \boldsymbol{r}_0 \\
&= \boldsymbol{x}_0 + \Sigma_0 A^\top \begin{bmatrix} S_{m-1} & \boldsymbol{s}_m \end{bmatrix} \begin{bmatrix} S_{m-1}^\top \\ \boldsymbol{s}_m^\top \end{bmatrix} \boldsymbol{r}_0 \\
&= \boldsymbol{x}_0 + \underbrace{\Sigma_0 A^\top S_{m-1} S_{m-1}^\top \boldsymbol{r}_0}_{=\boldsymbol{x}_{m-1}} + \Sigma_0 A^\top \boldsymbol{s}_m \boldsymbol{s}_m^\top \boldsymbol{r}_0.
\end{aligned}$$

It therefore remains to show that $\boldsymbol{s}_m^\top \boldsymbol{r}_0 = \boldsymbol{s}_m^\top \boldsymbol{r}_{m-1}$. To this end, from Eq. 5 we have

$$\begin{aligned}
\boldsymbol{s}_m^\top \boldsymbol{r}_{m-1} &= \boldsymbol{s}_m^\top \boldsymbol{b} - \boldsymbol{s}_m^\top A \boldsymbol{x}_{m-1} \\
&= \boldsymbol{s}_m^\top \boldsymbol{b} - \boldsymbol{s}_m^\top \boldsymbol{x}_0 - \underbrace{\boldsymbol{s}_m^\top A \Sigma_0 A^\top S_{m-1}^\top}_{=0} \boldsymbol{r}_0 \\
&= \boldsymbol{s}_m^\top \boldsymbol{r}_0
\end{aligned}$$

which completes the proof. □

**Lemma S3.** *Assume that the search directions $\{\boldsymbol{s}_i\}$ are $A\Sigma_0 A^\top$-orthogonal. At iteration $m$, the residual $\boldsymbol{r}_m = \boldsymbol{b} - A\boldsymbol{x}_m$ satisfies $\boldsymbol{r}_m^\top \boldsymbol{s}_i = 0$ for $i = 1, \ldots, m$.*

*Proof of Lemma S3.* By definition of $\boldsymbol{r}_m$ and $\boldsymbol{x}_m$

$$\begin{aligned}
\boldsymbol{s}_i^\top \boldsymbol{r}_m &= \boldsymbol{s}_i^\top \boldsymbol{b} - \boldsymbol{s}_i^\top A \boldsymbol{x}_m \\
&= \boldsymbol{s}_i^\top \boldsymbol{b} - \boldsymbol{s}_i^\top A \boldsymbol{x}_0 - \boldsymbol{s}_i^\top A \Sigma_0 A^\top S_m \Lambda_m^{-1} S_m^\top \boldsymbol{r}_0
\end{aligned}$$

Note that $\boldsymbol{s}_i^\top A\Sigma_0 A^\top S_m \Lambda_m^{-1} = \boldsymbol{e}_i^\top$, the vector with $[\boldsymbol{e}_i]_j = \delta_{ij}$, since $\boldsymbol{s}_i^\top A \Sigma_0 A^\top S_m$ is the $i^{\text{th}}$ row of $\Lambda_m$, whenever $i \leq m$. Thus, $\boldsymbol{s}_i^\top \boldsymbol{r}_m = \boldsymbol{s}_i^\top \boldsymbol{r}_0 - \boldsymbol{e}_i^\top S_m^\top \boldsymbol{r}_0 = 0$, as required. □

*Proof of Proposition 7.* Let $\tilde{\boldsymbol{t}}_1 := \boldsymbol{r}_0$, and for each $m > 1$, define $\tilde{\boldsymbol{t}}_m$ as

$$\tilde{\boldsymbol{t}}_m := \boldsymbol{r}_{m-1} - \sum_{i=1}^{m-1} \left( \boldsymbol{r}_{m-1}^\top Q \boldsymbol{t}_i \right) \boldsymbol{t}_i. \tag{S4}$$

where $Q = A\Sigma_0 A^\top$. Let $\boldsymbol{t}_m = \tilde{\boldsymbol{t}}_m / \|\tilde{\boldsymbol{t}}_m\|_Q$. We will show, inductively, that for each $m$ the set of search directions $\{\boldsymbol{t}_i\}_{i=1}^m$ is $Q$-orthonormal, and further that each $\boldsymbol{t}_i = \boldsymbol{s}_i$, as defined in the proposition statement.

For $m = 1$ the set $\{\boldsymbol{t}_1\}$ is trivially $Q$-orthonormal and $\boldsymbol{t}_1 = \boldsymbol{s}_1$. For $m > 1$ suppose $\{\boldsymbol{t}_i\}_{i=1}^{m-1}$ is $Q$-orthonormal and such that $\boldsymbol{t}_i = \boldsymbol{s}_i$, for $i = 1, \dots, m-1$. Then, for $j < m$

$$\begin{aligned}\boldsymbol{t}_j^\top Q\tilde{\boldsymbol{t}}_m &= \boldsymbol{t}_j^\top Q\boldsymbol{r}_{m-1} - \sum_{i=1}^{m-1} \boldsymbol{r}_{m-1}^\top Q\boldsymbol{t}_i \cdot \underbrace{\boldsymbol{t}_j^\top Q\boldsymbol{t}_i}_{=\delta_{ij}} \qquad \text{(by the inductive assumption)} \\ &= \boldsymbol{t}_j^\top Q\boldsymbol{r}_{m-1} - \boldsymbol{t}_j^\top Q\boldsymbol{r}_{m-1} = 0 \end{aligned} \tag{S5}$$

which shows that the set $\{\boldsymbol{t}_i\}_{i=1}^{m}$ is $Q$-orthonormal.

As a result we can apply Proposition 6 to show that

$$\begin{aligned} \boldsymbol{r}_j &= \boldsymbol{b} - A\boldsymbol{x}_j \\ &= \boldsymbol{b} - A\boldsymbol{x}_{j-1} - Q\boldsymbol{t}_j(\boldsymbol{t}_j^\top \boldsymbol{r}_{j-1}) \\ \implies Q\boldsymbol{t}_j &= \frac{\boldsymbol{r}_{j-1} - \boldsymbol{r}_j}{\boldsymbol{t}_j^\top \boldsymbol{r}_{j-1}} \\ \implies \boldsymbol{r}_{m-1}^\top Q\boldsymbol{t}_j &= \frac{\boldsymbol{r}_{m-1}^\top \boldsymbol{r}_{j-1} - \boldsymbol{r}_{m-1}^\top \boldsymbol{r}_j}{\boldsymbol{t}_j^\top \boldsymbol{r}_{j-1}}. \end{aligned} \tag{S6}$$

Since the set $\{\boldsymbol{t}_i\}_{i=1}^{m}$ is $Q$-orthonormal, we have from Lemma S3 that for each $j \leq m$, $\boldsymbol{r}_m^\top \boldsymbol{t}_j = 0$. Thus, from Eq. (S5) for each $j \leq m$:

$$0 = \boldsymbol{r}_m^\top \tilde{\boldsymbol{t}}_j := \boldsymbol{r}_m^\top \boldsymbol{r}_{j-1} - \sum_{i=1}^{m-1} \boldsymbol{r}_{m-1}^\top Q\boldsymbol{t}_i \cdot \underbrace{\boldsymbol{r}_m^\top \boldsymbol{t}_i}_{=0}. \tag{S7}$$

from which we conclude that $\boldsymbol{r}_m^\top \boldsymbol{r}_j = 0$ whenever $j < m$. It follows that Eq. (S6) is zero for all $j < m - 1$. Thus, all terms in the summation in Eq. (S4) vanish apart from the last, and we are left with

$$\tilde{\boldsymbol{t}}_m = \boldsymbol{r}_{m-1} - (\boldsymbol{r}_{m-1}^\top Q\boldsymbol{t}_{m-1})\boldsymbol{t}_{m-1}$$

which is equal to $\tilde{\boldsymbol{s}}_m$ for each $m > 1$, completing the proof. □

*Proof of Proposition 11.* First the posterior marginal for $\nu$ is computed. Note that

$$p(\nu|\boldsymbol{y}) \propto p(\boldsymbol{y}|\nu)p(\nu)$$

where

$$\begin{aligned} \boldsymbol{y}|\nu &\sim \mathcal{N}(S_m^\top A\boldsymbol{x}_0, \nu\Lambda_m) \\ \implies p(\nu|\boldsymbol{y}) &\propto \nu^{-\frac{m}{2}-1} \exp\left(-\frac{1}{2\nu}\boldsymbol{r}_0^\top S_m\Lambda_m^{-1}S_m^\top \boldsymbol{r}_0\right) \end{aligned}$$

which is $\mathsf{IG}\left(\frac{m}{2}, \frac{1}{2}\boldsymbol{r}_0^\top S_m\Lambda_m^{-1}S_m^\top \boldsymbol{r}_0\right)$. Now to determine the posterior marginal for $\boldsymbol{x}$

$$\begin{aligned} p(\boldsymbol{x}|\boldsymbol{y}) &= \int_0^\infty p(\boldsymbol{x}|\nu, \boldsymbol{y})p(\nu|\boldsymbol{y}) \,\mathrm{d}\nu \\ &\propto \int_0^\infty \nu^{-1-(m+d)/2} \exp\left(-\nu^{-1}K(\boldsymbol{x})\right) \,\mathrm{d}\nu \end{aligned} \tag{S8}$$

where

$$K(\boldsymbol{x}) := \frac{1}{2}\left[\boldsymbol{r}_0^\top S_m \Lambda_m^{-1} S_m^\top \boldsymbol{r}_0 + (\boldsymbol{x} - \boldsymbol{x}_m)^\top \Sigma_m^{-1} (\boldsymbol{x} - \boldsymbol{x}_m)\right]$$

Eq. S8 is recognised as the integral of an unnormalised inverse-Gamma density, so that

$$\begin{aligned} p(\boldsymbol{x}|\boldsymbol{y}) &\propto \Gamma(m+d) K(\boldsymbol{x})^{-\frac{1}{2}(m+d)} \\ &\propto \left[1 + \frac{1}{m}(\boldsymbol{x} - \boldsymbol{x}_m)^\top \left\{\frac{\boldsymbol{r}_0^\top S_m \Lambda_m^{-1} S_m^\top \boldsymbol{r}_0}{m} \Sigma_m\right\}^{-1} (\boldsymbol{x} - \boldsymbol{x}_m)\right]^{-\frac{1}{2}(m+d)} \end{aligned}$$

and therefore

$$p(\boldsymbol{x}|\boldsymbol{y}) = \mathsf{MVT}_m\left(\boldsymbol{x}_m, \frac{\boldsymbol{r}_0^\top S_m \Lambda_m^{-1} S_m^\top \boldsymbol{r}_0}{m} \Sigma_m\right)$$

□

**Proposition S4.** *It holds that* $\boldsymbol{x}_m \in \boldsymbol{x}_0 + K_{m-1}(\Sigma_0 A^\top A, \Sigma_0 A^\top \boldsymbol{r}_0)$.

*Proof of Proposition S4.* Let $\bar{K}_m = K_m(\Sigma_0 A^\top A, \Sigma_0 A^\top \boldsymbol{r}_0)$. Proof is by induction, with the additional inductive claims that

$$\Sigma_0 A^\top \boldsymbol{s}_m \in \bar{K}_{m-1} \tag{S9}$$

$$\Sigma_0 A^\top \boldsymbol{r}_m \in \bar{K}_m. \tag{S10}$$

Note that Eq. (S9) implies the required result by Proposition 1. Let $Q = A\Sigma_0 A^\top$. For $m = 1$, the first search direction is given by

$$\boldsymbol{s}_1 = \frac{\boldsymbol{r}_0}{\|\boldsymbol{r}_0\|_Q}$$

from which Eq. (S9) is clear. Further,

$$\begin{aligned} \boldsymbol{r}_1 &= \boldsymbol{b} - A\boldsymbol{x}_1 \\ &= \boldsymbol{b} - A\boldsymbol{x}_0 - \frac{A\Sigma_0 A^\top \boldsymbol{r}_0(\boldsymbol{r}_0^\top \boldsymbol{r}_0)}{\|\boldsymbol{r}_0\|_Q^2} \\ &= \boldsymbol{r}_0 - \frac{A\Sigma_0 A^\top \boldsymbol{r}_0(\boldsymbol{r}_0^\top \boldsymbol{r}_0)}{\|\boldsymbol{r}_0\|_Q^2} \\ \implies \Sigma_0 A^\top \boldsymbol{r}_1 &= \Sigma_0 A^\top \boldsymbol{r}_0 - \frac{(\Sigma_0 A^\top A)\Sigma_0 A^\top \boldsymbol{r}_0(\boldsymbol{r}_0^\top \boldsymbol{r}_0)}{\|\boldsymbol{r}_0\|_Q^2} \end{aligned}$$

from which it is clear that $\Sigma_0 A^\top \boldsymbol{r}_1 \in \bar{K}_1$.

Now for the inductive step. Assume that Equations (S9) and (S10) hold true up to $m-1$. From Proposition 7 we have that

$$\begin{aligned} \tilde{\boldsymbol{s}}_m &= \boldsymbol{r}_{m-1} - (\boldsymbol{r}_{m-1}^\top Q \boldsymbol{s}_{m-1})\boldsymbol{s}_{m-1} \\ \implies \Sigma_0 A^\top \tilde{\boldsymbol{s}}_m &= \underbrace{\Sigma_0 A^\top \boldsymbol{r}_{m-1}}_{\in \bar{K}_{m-1}} - (\boldsymbol{r}_{m-1}^\top Q \boldsymbol{s}_{m-1}) \underbrace{\Sigma_0 A^\top \boldsymbol{s}_{m-1}}_{\in \bar{K}_{m-2}} \end{aligned}$$

where inclusion in the Krylov subspaces is by the inductive assumption. It follows that $\Sigma_0 A^\top \boldsymbol{s}_m \in \bar{K}_{m-1}$. Lastly, observe that

$$\begin{aligned}
\boldsymbol{r}_m &= \boldsymbol{b} - A\boldsymbol{x}_m \\
&= \boldsymbol{r}_{m-1} - A\Sigma_0 A^\top \boldsymbol{s}_m(\boldsymbol{s}_m^\top \boldsymbol{r}_m) \\
\implies \Sigma_0 A^\top \boldsymbol{r}_m &= \underbrace{\Sigma_0 A^\top \boldsymbol{r}_{m-1}}_{\in \bar{K}_{m-1}} - \underbrace{(\Sigma_0 A^\top A)\Sigma_0 A^\top \boldsymbol{s}_m}_{\in \bar{K}_m}(\boldsymbol{s}_m^\top \boldsymbol{r}_m)
\end{aligned}$$

which by the inductive assumption is in $\bar{K}_m$, as required. □

*Proof of Proposition 9.* Let $Q = A\Sigma_0 A^\top$. Begin with $m = 1$. Any $\boldsymbol{x} \in K_0^*$ can be represented as $\boldsymbol{x}_0 + \alpha_1 \Sigma_0 A^\top \boldsymbol{r}_0$ for some $\alpha_1$. Thus, when $\boldsymbol{x} \in K_1^*$:

$$\begin{aligned}
\|\boldsymbol{x} - \boldsymbol{x}^*\|_{\Sigma_0^{-1}}^2 &= \|\boldsymbol{x}_0 + \alpha_1 \Sigma_0 A^\top \boldsymbol{r}_0 - \boldsymbol{x}^*\|_{\Sigma_0^{-1}}^2 \\
&= \boldsymbol{x}_0^\top \Sigma_0^{-1} \boldsymbol{x}_0 + 2\alpha_1 \boldsymbol{x}_0^\top A^\top \boldsymbol{r}_0 - 2\boldsymbol{x}_0^\top \Sigma_0^{-1} \boldsymbol{x}^* \\
&\quad + \alpha_1^2 \boldsymbol{r}_0^\top A\Sigma_0 A^\top \boldsymbol{r}_0 - 2\alpha_1 \boldsymbol{r}_0^\top A\boldsymbol{x}^* \\
&\quad + (\boldsymbol{x}^*)^\top \Sigma_0^{-1} \boldsymbol{x}^* \\
\implies \frac{\mathrm{d}}{\mathrm{d}\alpha_1} \|\boldsymbol{x} - \boldsymbol{x}^*\|_{\Sigma_0^{-1}}^2 &= 2\boldsymbol{x}_0^\top A^\top \boldsymbol{r}_0 + 2\alpha_1 \boldsymbol{r}_0^\top A\Sigma_0 A^\top \boldsymbol{r}_0 - 2\boldsymbol{r}_0^\top A\boldsymbol{x}^*.
\end{aligned}$$

Setting this to zero, we obtain:

$$\alpha_1 = \frac{\boldsymbol{r}_0^\top (\boldsymbol{b} - A\boldsymbol{x}_0)}{\|\boldsymbol{r}_0\|_Q^2} = \frac{\boldsymbol{r}_0^\top \boldsymbol{r}_0}{\|\boldsymbol{r}_0\|_Q^2}.$$

From Proposition 6, this corresponds to $\boldsymbol{x} = \boldsymbol{x}_1$. It is further clear that

$$\frac{\mathrm{d}^2}{\mathrm{d}\alpha_1^2} \|\boldsymbol{x} - \boldsymbol{x}^*\|_{\Sigma_0^{-1}}^2 = 2\|\boldsymbol{r}_0\|_Q^2 > 0$$

so that $\boldsymbol{x}_1$ is optimal in $K_0^*$.

Now observe that $\Sigma_0 A^\top \boldsymbol{s}_m$ is orthogonal to $\boldsymbol{x}_{m-1} - \boldsymbol{x}_0$ in the $\Sigma_0^{-1}$-inner-product:

$$\begin{aligned}
\left\langle \Sigma_0 A^\top \boldsymbol{s}_m, \boldsymbol{x}_{m-1} - \boldsymbol{x}_0 \right\rangle_{\Sigma_0^{-1}} &= \boldsymbol{s}_m^\top A\boldsymbol{x}_0 + \underbrace{\boldsymbol{s}_m^\top A\Sigma_0 A^\top S_{m-1}}_{=0}(S_{m-1}^\top \boldsymbol{r}_0) - \boldsymbol{s}_m^\top A\boldsymbol{x}_0 \\
&= 0
\end{aligned}$$

As a result, for $m > 1$ it suffices to determine $\alpha_m$ in

$$\begin{aligned}
\boldsymbol{x} &= \boldsymbol{x}_0 + (\boldsymbol{x}_{m-1} - \boldsymbol{x}_0) + \alpha_m \Sigma_0 A^\top \boldsymbol{s}_m \\
&= \boldsymbol{x}_{m-1} + \alpha_m \Sigma_0 A^\top \boldsymbol{s}_m
\end{aligned}$$

where $\boldsymbol{x} \in K^*_{m-1}$. Again, $\alpha_m$ is determined directly, much as above:

$$\begin{aligned}
\|\boldsymbol{x} - \boldsymbol{x}^*\|^2_{\Sigma_0^{-1}} &= \|\boldsymbol{x}_{m-1} + \alpha_m \Sigma_0 A^\top \boldsymbol{s}_m - \boldsymbol{x}^*\|^2_{\Sigma_0^{-1}} \\
\implies \frac{\mathrm{d}}{\mathrm{d}\alpha_m} \|\boldsymbol{x} - \boldsymbol{x}^*\|^2_{\Sigma_0^{-1}} &= 2\boldsymbol{x}_{m-1}^\top A^\top \boldsymbol{s}_m + 2\alpha_m \boldsymbol{s}_m^\top A \Sigma_0 A^\top \boldsymbol{s}_m - 2\boldsymbol{s}_m^\top A \boldsymbol{x}^*. \\
\implies \alpha_m &= \frac{\boldsymbol{s}_m^\top (\boldsymbol{b} - A\boldsymbol{x}_{m-1})}{\|\boldsymbol{s}_m\|^2_Q} = \boldsymbol{s}_m^\top \boldsymbol{r}_{m-1}
\end{aligned}$$

which is also a minimum. Thus, we have that

$$\underset{\boldsymbol{x} \in K^*_{m-1}}{\arg\min} \|\boldsymbol{x} - \boldsymbol{x}^*\|^2_{\Sigma_0^{-1}} = \boldsymbol{x}_{m-1} + \Sigma_0 A^\top \boldsymbol{s}_m (\boldsymbol{s}_m^\top \boldsymbol{r}_m) \equiv \boldsymbol{x}_m$$

from Proposition 6, which completes the proof. □

*Proof of Proposition 10.* We begin by introducing the *operator norm* induced by the energy norm $\|\cdot\|_A$, which is a norm on matrices $M \in \mathbb{R}^{d \times d}$

$$\|M\|_A^{\text{op}} = \sup \left\{ \|M\boldsymbol{v}\|_A : \|\boldsymbol{v}\|_A = 1 \right\}.$$

From Proposition S4 it holds that there exists a polynomial $\tilde{P}_{m-1}$ of degree $m-1$ such that

$$\begin{aligned}
\boldsymbol{e}_m := \boldsymbol{x}_m - \boldsymbol{x}^* &= \boldsymbol{x}_0 - \boldsymbol{x}^* + \tilde{P}_{m-1}(\Sigma_0 A^\top A) \Sigma_0 A^\top \boldsymbol{r}_0 \\
&= \boldsymbol{e}_0 + \tilde{P}_{m-1}(\Sigma_0 A^\top A) \Sigma_0 A^\top A \boldsymbol{e}_0 \\
&= P_m(\Sigma_0 A^\top A) \boldsymbol{e}_0
\end{aligned}$$

where $P_m$ is some polynomial of degree $m$. Thus

$$\begin{aligned}
\|\boldsymbol{e}_m\|_{\Sigma_0^{-1}} &\leq \|P_m(\Sigma_0 A^\top A)\|^{\text{op}}_{\Sigma_0^{-1}} \cdot \|\boldsymbol{e}_0\|_{\Sigma_0^{-1}} \\
&= \|\Sigma_0^{-\frac{1}{2}} P_m(\Sigma_0 A^\top A) \Sigma_0^{\frac{1}{2}}\|_I^{\text{op}} \cdot \|\boldsymbol{e}_0\|_{\Sigma_0^{-1}} \\
&= \|P_m(\Sigma_0^{\frac{1}{2}} A^\top A \Sigma_0^{\frac{1}{2}})\|_I^{\text{op}} \cdot \|\boldsymbol{e}_0\|_{\Sigma_0^{-1}}
\end{aligned}$$

Now, note that $\Sigma_0^{\frac{1}{2}} A^\top A \Sigma_0^{\frac{1}{2}}$ is symmetric, and can thus be represented as $\Sigma_0^{\frac{1}{2}} A^\top A \Sigma_0^{\frac{1}{2}} = V \Gamma V^\top$, where $\Gamma$ is the matrix with the eigenvalues of $\Sigma_0^{\frac{1}{2}} A^\top A \Sigma_0^{\frac{1}{2}}$ on its diagonal, and $V$ is the orthonormal matrix of its eigenvectors. Furthermore note that $\Sigma_0 A^\top A = \Sigma_0^{\frac{1}{2}} [\Sigma_0^{\frac{1}{2}} A^\top A \Sigma_0^{\frac{1}{2}}] \Sigma_0^{-\frac{1}{2}}$. Hence, $\Sigma_0 A^\top A$ is similar to $\Sigma_0^{\frac{1}{2}} A^\top A \Sigma_0^{\frac{1}{2}}$, and so the matrices share the same eigenvalues.

Now, clearly $P_m(V \Gamma V^\top) = V P_m(\Gamma) V^\top$ since $V$ is orthonormal. Thus

$$\begin{aligned}
\|\boldsymbol{e}_m\|_I &\leq \underbrace{\|V\|_I^{\text{op}} \|V^\top\|_I^{\text{op}}}_{=1} \|P_m(\Gamma)\|_I^{\text{op}} \cdot \|\boldsymbol{e}_0\|_{\Sigma_0^{-1}} \\
&= \|P_m(\Gamma)\|_I^{\text{op}} \cdot \|\boldsymbol{e}_0\|_{\Sigma_0^{-1}} \qquad \text{(S11)}
\end{aligned}$$

where $\|V\|_I^{\text{op}}\|V^\top\|_I^{\text{op}} = 1$ follows since $V$ is unitary. Let $\mathbb{P}_m$ denote the set of all polynomials of order $m$ with the property that $P(0) = 1$ for each $P \in \mathbb{P}_m$. This requirement ensures that if $A$ is singular, $\|\boldsymbol{e}_m\|_{\Sigma_0^{-1}} = \|\boldsymbol{e}_0\|_{\Sigma_0^{-1}}$ for all $m$. Now, from Proposition 9 we have that $P_m \in \mathbb{P}_m$ is constructed to minimise the error $\boldsymbol{e}_m$. Let $\bar{\Gamma}$ denote the set of eigenvalues of $\Sigma_0^{\frac{1}{2}} A^\top A \Sigma_0^{\frac{1}{2}}$. Then

$$\begin{aligned}
\|P_m(\Gamma)\|_I^{\text{op}} &= \min_{P\in\mathbb{P}_m} \max_{\gamma\in\bar{\Gamma}} \sup_{\|\boldsymbol{v}\|_2=1} \|P(\gamma)\boldsymbol{v}\|_2 \\
&= \min_{P\in\mathbb{P}_m} \max_{\gamma\in\bar{\Gamma}} |P(\gamma)| \\
&\leq \min_{P\in\mathbb{P}_m} \max_{\gamma\in[\gamma_{\min},\gamma_{\max}]} |P(\gamma)| \qquad \text{(S12)}
\end{aligned}$$

Lemma S5, proven below, establishes that the polynomial minimising this expression is

$$P(\gamma) = \frac{T_m\left(\frac{\gamma_{\max}+\gamma_{\min}-2\gamma}{\gamma_{\max}-\gamma_{\min}}\right)}{T_m\left(\frac{\gamma_{\max}+\gamma_{\min}}{\gamma_{\max}-\gamma_{\min}}\right)}$$

where $T_m(\cdot)$ is the $m^{\text{th}}$ Chebyshev polynomial of the first kind.

Let $\kappa = \gamma_{\max}/\gamma_{\min}$. Now, $T_m(z) \in [-1, 1]$ for all $m$ and all $z \in [-1, 1]$; thus the numerator takes maximum value 1. Therefore

$$\|P_m(\Gamma)\|_{\Sigma_0^{-1}}^{\text{op}} \leq \left|T_m\left(\frac{\kappa+1}{\kappa-1}\right)\right|^{-1}.$$

Lastly, note that by definition

$$T_m(z) = \frac{1}{2}\left[\left(z + \sqrt{z^2-1}\right)^m + \left(z - \sqrt{z^2-1}\right)^m\right]$$

so that

$$\begin{aligned}
\|P_m(\Gamma)\|_2^{\text{op}} &\leq 2\left[\left(\frac{\sqrt{\kappa}+1}{\sqrt{\kappa}-1}\right)^m + \left(\frac{\sqrt{\kappa}-1}{\sqrt{\kappa}+1}\right)^m\right]^{-1} \\
&\leq 2\left(\frac{\sqrt{\kappa}-1}{\sqrt{\kappa}+1}\right)^m.
\end{aligned}$$

Inserting this into Eq. (S11) and recalling that since $\Sigma_0^{\frac{1}{2}} A^\top A \Sigma_0^{\frac{1}{2}}$ has the same eigenvalues as $\Sigma_0 A^\top A$, it also has the same condition number, completes the proof. □

**Lemma S5** (Appendix S3 of Shewchuk [1994])**.** *Eq.* (S12) *is minimised by*

$$P(\gamma) = \frac{T_m\left(\frac{\gamma_{\max}+\gamma_{\min}-2\gamma}{\gamma_{\max}-\gamma_{\min}}\right)}{T_m\left(\frac{\gamma_{\max}+\gamma_{\min}}{\gamma_{\max}-\gamma_{\min}}\right)}$$

*where $T_m$ is the $m^{th}$ Chebyshev polynomial of the first kind.*

*Proof.* For convenience let

$$\gamma_0 := \frac{\gamma_{\max} + \gamma_{\min}}{\gamma_{\max} - \gamma_{\min}}$$

and note that $\gamma_0 > 1$. Further, observe that

$$\gamma \in [\gamma_{\min}, \gamma_{\max}] \implies \frac{\gamma_{\max} + \gamma_{\min} - 2\gamma}{\gamma_{\max} - \gamma_{\min}} \in [-1, 1].$$

Now recall the following properties of Chebyshev polynomials:

**C1** $T_m(z) \in [-1, 1]$ for all $z \in [-1, 1]$.

**C2** $T_m(1) = 1$, and $T_m(-1) = (-1)^m$.

**C3** Let $Z = \{z_i\}, i = 1, \dots, m$ denote the ordered zeros of $T_m(z)$. Then, $Z \subset [-1, 1]$.

**C4** $T_m(z)$ attains the value $(-1)^{m+i}$ in the range $[z_i, z_{i+1}]$ for $i = 1, \dots, m-1$.

First, note that clearly $P(0) = 1$ as $T_m(\gamma_0) \neq 0$. This is because $\gamma_0 > 1$ and so $T_m(\gamma_0) > 1$ by **C2** and **C3**. Thus, $P(\gamma) \in \mathbb{P}_m$ as required. Further, note that

$$\max_{\gamma \in [\gamma_{\min}, \gamma_{\max}]} |P(\gamma)| = T_m(\gamma_0)^{-1}$$

by **C1**. Proof that $P(\gamma)$ minimizes Eq. (S12) is by contradiction. Suppose there is a $Q(\gamma) \in \mathbb{P}_m$ with

$$\max_{\gamma \in [\gamma_{\min}, \gamma_{\max}]} |Q(\gamma)| < T_m(\gamma_0)^{-1} \tag{S13}$$

Consider the polynomial $P(\gamma) - Q(\gamma)$. From **C1**, $P(\gamma) \in [-T_m(\gamma_0)^{-1}, T_m(\gamma_0)^{-1}]$, and $P(\gamma)$ has $m$ zeros in $[\gamma_{\min}, \gamma_{\max}]$. From Eq. S13 it is clear that $P(\gamma) - Q(\gamma)$ also has $m$ zeros in $[\gamma_{\min}, \gamma_{\max}]$, as to prevent $P(\gamma)$ from crossing zero between its extrema in this range would require $|Q(\gamma)| > T_m(\gamma_0)^{-1}$ (by **C4**).

However, since $P(0) = Q(0) = 1$, $P - Q$ has an additional zero outside $[\gamma_{\min}, \gamma_{\max}]$. Therefore, $P - Q$ is a polynomial of degree $m$ with at least $m + 1$ zeros, which is a contradiction. Thus $P(\gamma)$ minimises Eq. S12.

□

# S2 Further Simplication of BayesCG

In this section, the simplifications mentioned in Section 5 and exploited in Algorithm 1 are described in detail. First, two coefficients must be calculated, one to update $\boldsymbol{x}_m$ and one to update $\tilde{\boldsymbol{s}}_m$. Note that for stability reasons we work with un-normalized rather than normalized search directions where possible. As usual let $Q = A\Sigma_0 A^\top$, and express these quantities as

$$\boldsymbol{x}_m = \boldsymbol{x}_{m-1} + \alpha_m \Sigma_0 A^\top \tilde{\boldsymbol{s}}_m$$
$$\tilde{\boldsymbol{s}}_m = \boldsymbol{r}_{m-1} + \beta_{m-1} \tilde{\boldsymbol{s}}_{m-1}$$

where

$$\alpha_m = \frac{\tilde{\boldsymbol{s}}_m^\top \boldsymbol{r}_{m-1}}{\|\tilde{\boldsymbol{s}}_m\|_Q^2}$$

$$\beta_m = -\frac{\boldsymbol{r}_m Q \tilde{\boldsymbol{s}}_m}{\|\tilde{\boldsymbol{s}}_m\|_Q^2}$$

Now, using the expression for $\tilde{\boldsymbol{s}}_m$, note that

$$\begin{aligned}\alpha_m &= \frac{\boldsymbol{r}_{m-1}^\top(\boldsymbol{r}_{m-1} - \beta_m \tilde{\boldsymbol{s}}_{m-1})}{\|\tilde{\boldsymbol{s}}_m\|_Q^2} \\ &= \frac{\boldsymbol{r}_{m-1}^\top \boldsymbol{r}_{m-1}}{\|\tilde{\boldsymbol{s}}_m\|_Q^2}\end{aligned}$$

since, from Lemma S3, $\tilde{\boldsymbol{s}}_m^\top \boldsymbol{r}_m = 0$. Furthermore, from the proof of Proposition 7, we have

$$\begin{aligned}\boldsymbol{r}_m^\top Q \boldsymbol{s}_m &= \frac{\boldsymbol{r}_m^\top \boldsymbol{r}_{m-1} - \boldsymbol{r}_m^\top \boldsymbol{r}_m}{\boldsymbol{s}_m^\top \boldsymbol{r}_{m-1}} \\ &= -\frac{\boldsymbol{r}_m^\top \boldsymbol{r}_m}{\boldsymbol{s}_m^\top \boldsymbol{r}_{m-1}} \\ &= -\frac{\boldsymbol{r}_m^\top \boldsymbol{r}_m}{\boldsymbol{r}_{m-1}^\top \boldsymbol{r}_{m-1}} \|\tilde{\boldsymbol{s}}_m\|_Q^2\end{aligned}$$

so that

$$\beta_m = \frac{\boldsymbol{r}_m^\top \boldsymbol{r}_m}{\boldsymbol{r}_{m-1}^\top \boldsymbol{r}_{m-1}}$$

These two simplifications allow rearranging the expressions in Proposition 6 into Algorithm 1.

# S3 Additional Numerical Results for Simulation Study

In this section we discuss an empirical procedure for calibrating the scale of the posterior covariance, in an attempt to compensate for the fact that search directions depend upon $\boldsymbol{x}^*$, and show that this results in better calibrated UQ. The proposed approach is to construct an error indicator over the course of the algorithm, and then use this to adjust an appropriate measure of spread of the posterior to match that error prediction.

**1. Constructing the Error Indicator** The aim here is to construct a proxy for the true error by measuring the convergence of the BayesCG mean. Let

$$z_i := \|\boldsymbol{x}_i - \boldsymbol{x}_{i-1}\|_2 \,.$$

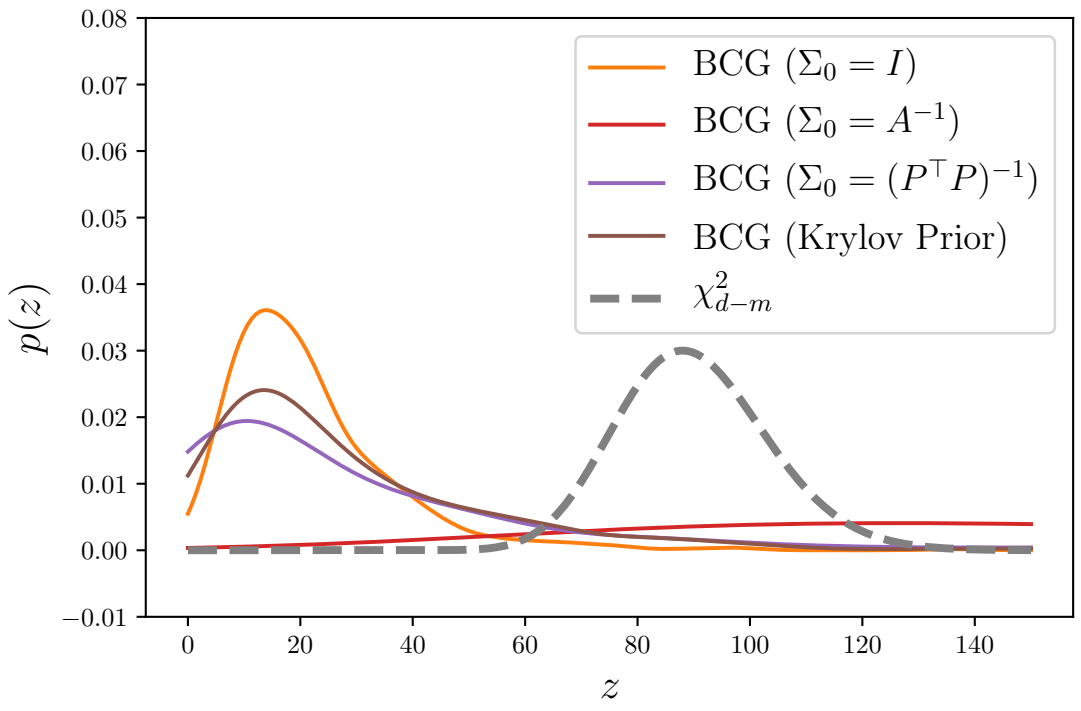

Figure S1: Uncertainty quantification provided in the new proposal. This should be compared against Fig. 3 in the main text.

The idea is to perform a simple regression on the values $\{z_i\}_{i=1}^m$ and use the fitted model $\nu(i)$ to extrapolate the error forward. Justified by the exponential convergence rate of BayesCG, as well as its simplicity, a log-linear function $\nu(i) = \exp(a + bi)$ has been used.

To derive our error indicator we use the following triangle inequality bound:

$$\begin{aligned} \|\boldsymbol{x}_m - \boldsymbol{x}^*\|_2 &\leq \sum_{i=m+1}^{d} \|\boldsymbol{x}_i - \boldsymbol{x}_{i-1}\|_2 \\ &\approx \sum_{i=m+1}^{d} \nu(i) =: \alpha_m. \end{aligned}$$

Thus $\alpha_m$ provides an approximate upper-bound for $\|\boldsymbol{x}_m - \boldsymbol{x}^*\|_2$.

**2. Fitting the Posterior** Next we adjust the spread of the posterior, in a somewhat *ad-hoc* manner, based on the approximate upper-bound $\alpha_m$ on the true error. This requires the posterior spread to be quantified, and for the ease of computability we used $\mathrm{trace}(\nu_m \Sigma_m)$. Thus, to be concrete, we would like to select $\nu_m$ so that the spread

$$\begin{aligned} \mathrm{trace}(\nu_m \Sigma_m) &= \alpha_m \\ \implies \nu_m &= \frac{\alpha_m}{\mathrm{trace}(\Sigma_m)}. \end{aligned}$$

Note that, since $\alpha_m$ appears in the numerator and provides an approximate upper bound for the true error, the UQ provided will still be conservative in general.

**3. Results** The UQ computed in the main text, Fig. 3, can be compared with the UQ under this proposal shown in Fig. S1. The UQ under this proposal is substantially better calibrated than when the Jeffrey's prior is used in all cases apart from the case when $\Sigma_0 = A^{-1}$. However since this performs well for all the practical choices of prior covariance suggested, the results indicate that approaches based on heuristic calibration of posterior spread could be used to compensate for the fact that the search directions have been constructed in a data-driven manner.

## S4 Experimental Results for Higher-Dimensional Systems

In this section additional experimental results are reported for higher-dimensional systems than those considered in Section 6.1. The experimental protocol adopted in that section is challenging to adapt to higher-dimensional systems, as the method for generating sparse positive-definite matrices has empirically been found to produce numerically singular matrices when $d$ is increased. As a result, in this section we will adopt a more structured approach to generating random systems, based on discretisation of a simple elliptic PDE.

Specifically, the PDE considered is the following PDE with random boundary conditions:

$$\begin{aligned} -\nabla u(\boldsymbol{z}) &= 0 && \boldsymbol{z} \in (0,1)^2 \\ \frac{\partial u}{\partial z_2}(\boldsymbol{z}) &= 0 && z_1 \in (0,1),\ z_2 \in \{0,1\} \\ u(\boldsymbol{z}) &= f(\boldsymbol{z}) && z_1 \in \{0,1\},\ z_2 \in [0,1] \end{aligned} \tag{S14}$$

where $\log f(\boldsymbol{z}) \sim \mathcal{GP}(0, k(\boldsymbol{z}, \boldsymbol{z}'; \rho))$ and $k(\boldsymbol{z}, \boldsymbol{z}'; \rho)$ is a Matérn covariance function:

$$k(\boldsymbol{z}, \boldsymbol{z}'; \rho) = \left(1 + \frac{\sqrt{3}}{\rho}\right) \exp\left(-\frac{\sqrt{3}\|\boldsymbol{z} - \boldsymbol{z}'\|_2}{\rho}\right).$$

For the purposes of this experiment the length-scale was fixed to $\rho = 0.1$.

Eq. (S14) was discretised with the finite-element method using standard piecewise linear basis functions as implemented in `FEniCS`, as in Section 6.2. The domain was discretised using a simple regular triangular mesh over the domain resulting in $d$ elements. To investigate the performance of BayesCG as a function of the dimension of the system, three different discretisation levels were used: $d = 121$, $d = 1089$ and $d = 10201$.

A subset of the priors from Section 6.1 were considered. The prior $\Sigma_0 = A^{-1}$ was not used, as for $d = 10201$ computing this is impractical. Similarly, the procedure we have used for calculating the Krylov subspace prior in that section requires knowledge of $\kappa(A)$, which is also impractical. As a result we have focussed on the prior $\Sigma_0 = (P^\top P)^{-1}$ where $P$ is, as in Section 6.1, a preconditioner based on an incomplete Cholesky factorisation of $A$. Results for $\Sigma_0 = I$ are also included.

Results for convergence of the posterior mean are reported in Fig. S2. Note that the number of iterations required for both CG and BCG with the preconditioner prior seems to increase sub-linearly with dimension, while with $\Sigma_0 = I$ there is qualitatively little difference as a function of dimension. Furthermore note that the rate of convergence of CG appears to overtake that of the preconditioner prior as the dimension increases, suggesting that the quality of the incomplete Cholesky preconditioner decays with dimension. However, we note that this is only one choice of preconditioner, and others preconditioners may behave better.

The quality of the UQ as a function of dimension is displayed in Fig. S3, with the statistic $Z$ as described in Section 6.1. The UQ after $\lfloor \frac{d}{10} \rfloor$ iterations was considered. The quality of UQ provided is seen to decrease as a function of dimension $d$.

## S5 Experimental Set-Up for EIT

The results presented in this paper used experimental data provided by EIDORS[1] and due to Isaacson et al. [2004]. In the experiment, depicted in Figure 8a, three targets were placed into a tank filled with saline, two of which are lung-shaped and one of which is heart-shaped. The lung-shaped targets have lower conductivity than the surrounding saline, while the heart-shaped target has higher conductivity. A total of 32 electrodes were placed around the boundary of the domain, and stimulated with 31 distinct stimulation patterns as described in Isaacson et al. [2004]. For each stimulation, the voltage induced at every electrode was recorded, and there are thus $32 \times 31$ distinct measurements on which the prior must be conditioned. The inducing currents and measured voltages were each supplied in the referenced dataset.

In the simulations the circular tank was modelled as a unit circular domain, and the electrodes were assumed to occupy precisely $1/64^{\text{th}}$ of the boundary. Thus, each electrode had length $\pi/32$ and there was a distance of $\pi/32$ between each neighbouring pair of electrodes on the boundary. Since no information is known on the quality of the electrode contact, we set the contact impedances to an arbitrary value, $\zeta_l = 1$ for each $l$.

The trangulations required to discretise the PDE were generated using the Python package `meshpy`, configured to ensure that there were $N_d$ equally sized elements on the boundary. $N_d$ was chosen to be a multiple of the number of boundary electrodes, so that each electrode corresponds to the same number of boundary elements, and other boundary elements are disjoint from all electrodes. Figure S4 shows an illustration of a triangulation of the domain used to discretise the PDE, with $N_d = 64$.

## S6 Additional Numerical Results for EIT

Figure S5 shows the posterior distribution obtained from BayesCG for different values of $\epsilon$. The linear system solved was generated for $N_d = 128$ and with the conductivity field $\hat{\sigma}(\boldsymbol{z})$. Plotted is the posterior mean from BayesCG, along with samples from the posterior distribution, over the spatial domain of the PDE. That is, the voltage field $v(\boldsymbol{z})$ has been plotted rather than the conductivity field from the inverse problem. The top row has the largest value of $\epsilon$, and here clearly the posterior mean deviates far from the true solution, depicted in the bottom row. However by $\epsilon = 5$ the mean from BayesCG appears close to the truth. The second, third and fourth column show samples from the posterior distribution, and while there is significantly more noise in these columns the main characteristics of the true solution are visible even at $\epsilon = 20$, suggesting that the use of BayesCG within a Bayesian approach to EIT can be qualitatively justified.

[1] At time of writing this data can be found at the EIDORS website.

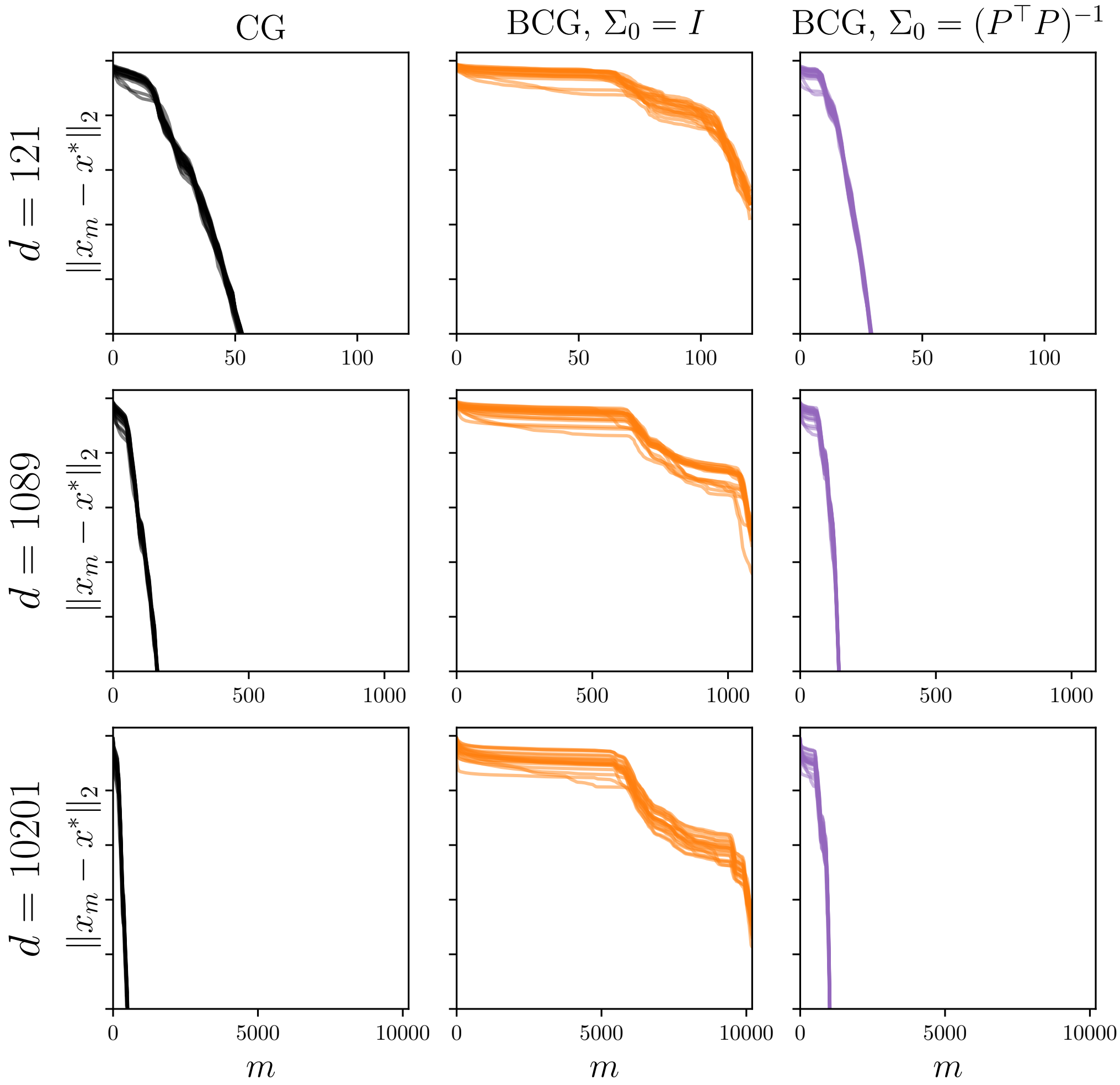

Figure S2: Convergence of the posterior mean from BayesCG, for the PDE example from Section S4, as dimension is varied. The error $\|\boldsymbol{x}_m = \boldsymbol{x}^*\|_2$ is reported for CG (left) as well as BayesCG with different prior covariances $\Sigma_0$.

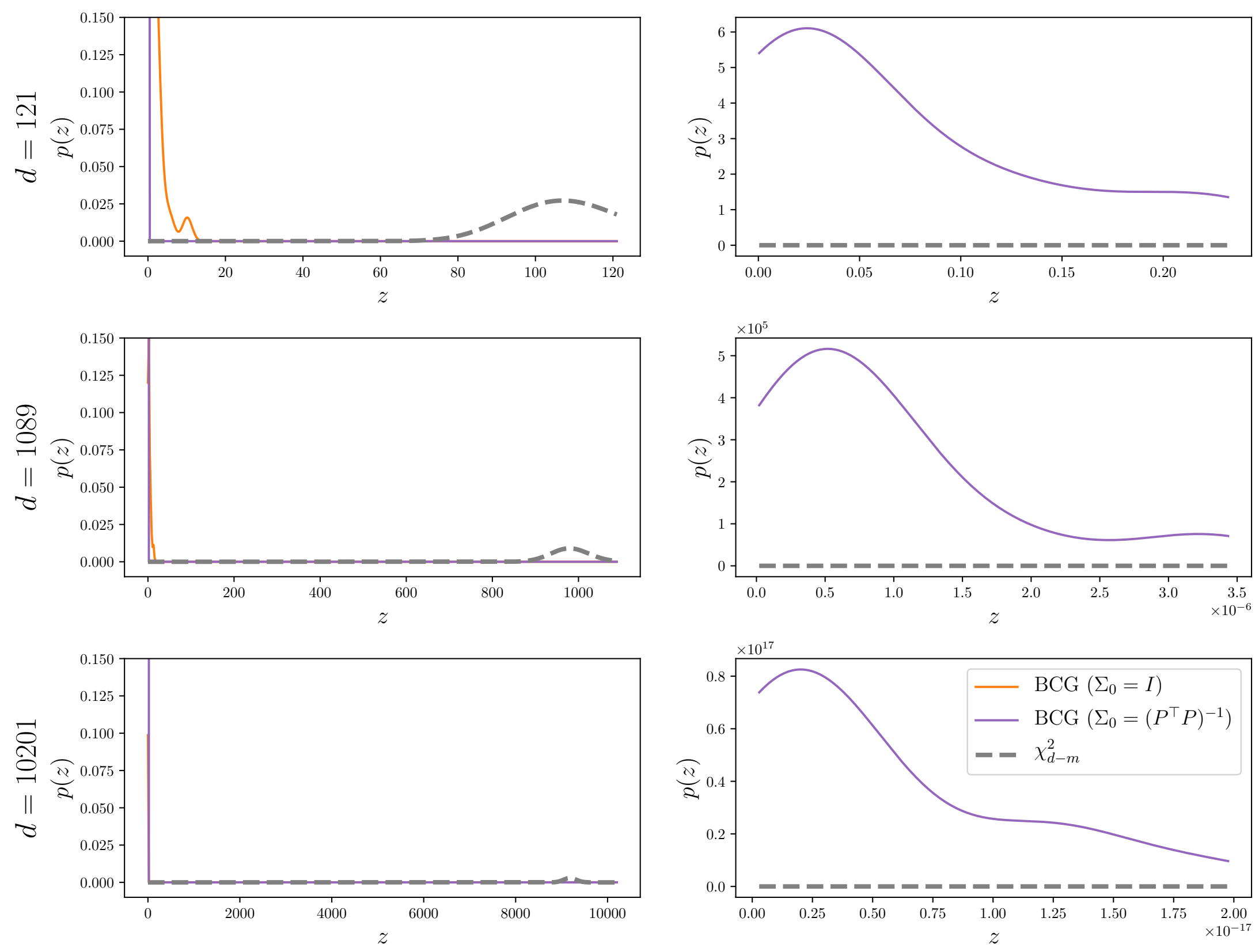

Figure S3: Assessment of the uncertainty quantification provided by the Gaussian version of BayesCG, for the PDE example from Section S4, as dimension is varied. The statistic $Z$ is reported based on 100 independently sampled test problems, for different prior covariances $\Sigma_0$. These are compared to the theoretical distribution of $Z$ when the posterior distribution is well-calibrated. The right panel zooms in on the portion of the $x$-axis occupied by the statistic for $\Sigma_0 = (P^\top P)^{-1}$.

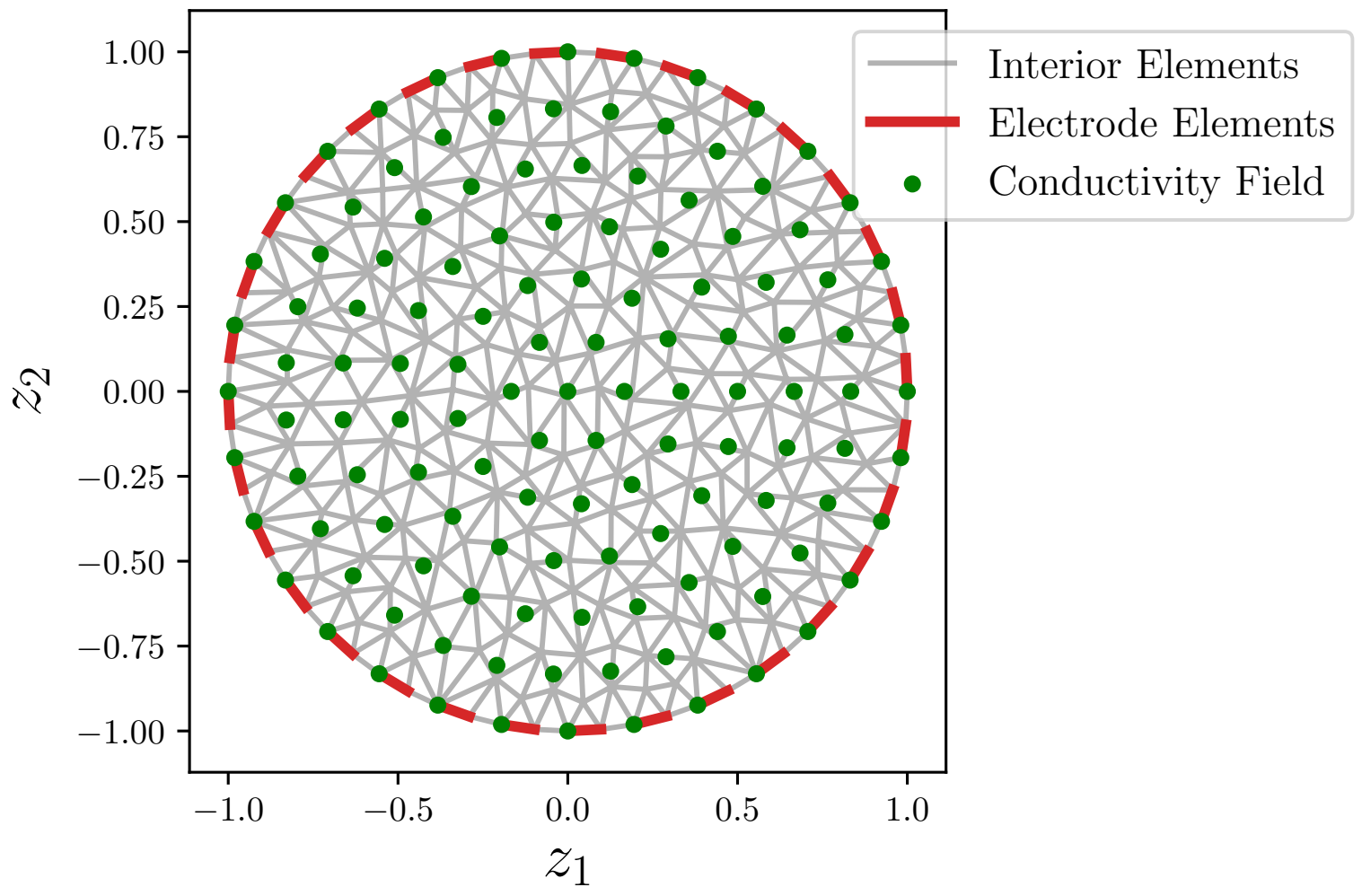

Figure S4: Finite-element discretisation used for the EIT experiment described in Section 6.2. The mesh was generated using the Python package `meshpy`, configured to ensure that there were 64 equally spaced boundary elements. Red lines indicate the elements which correspond to electrodes. Green dots show the locations at which the posterior conductivity field was sampled.

Fig. S6 shows the behavior of the posterior distribution when a standard CG forward solver is used but with $m$ held fixed. At $m = 40$ the computed mean bears no resemblance to the actual posterior mean. Moreover, the computed distribution is over-confident, as reflected by the uniformly *lower* computed standard deviation, compared to the actual posterior. At $m = 60$ the qualitative features of the posterior mean have been recovered, though the recovery still differs noticeably from that of the actual posterior, and the computed standard deviation remains lower. This provides further motivation for the use of BayesCG, as a means to constrain the solver to fewer iterations while still obtaining estimates that are statistically meaningful.

Fig. S7 repeats this experiment for the BayesCG forward solver when the preconditioner prior is used. Again, as $m$ is increased the posterior mean exhibits clear structure in the conductivity field. While the posterior variance does not visibly appear to decrease in the bottom row, the integrated standard deviation is nevertheless decreasing, starting at 0.0586 at $m = 40$, decreasing to 0.0459 at $m = 60$ and 0.0365 at $m = 80$.

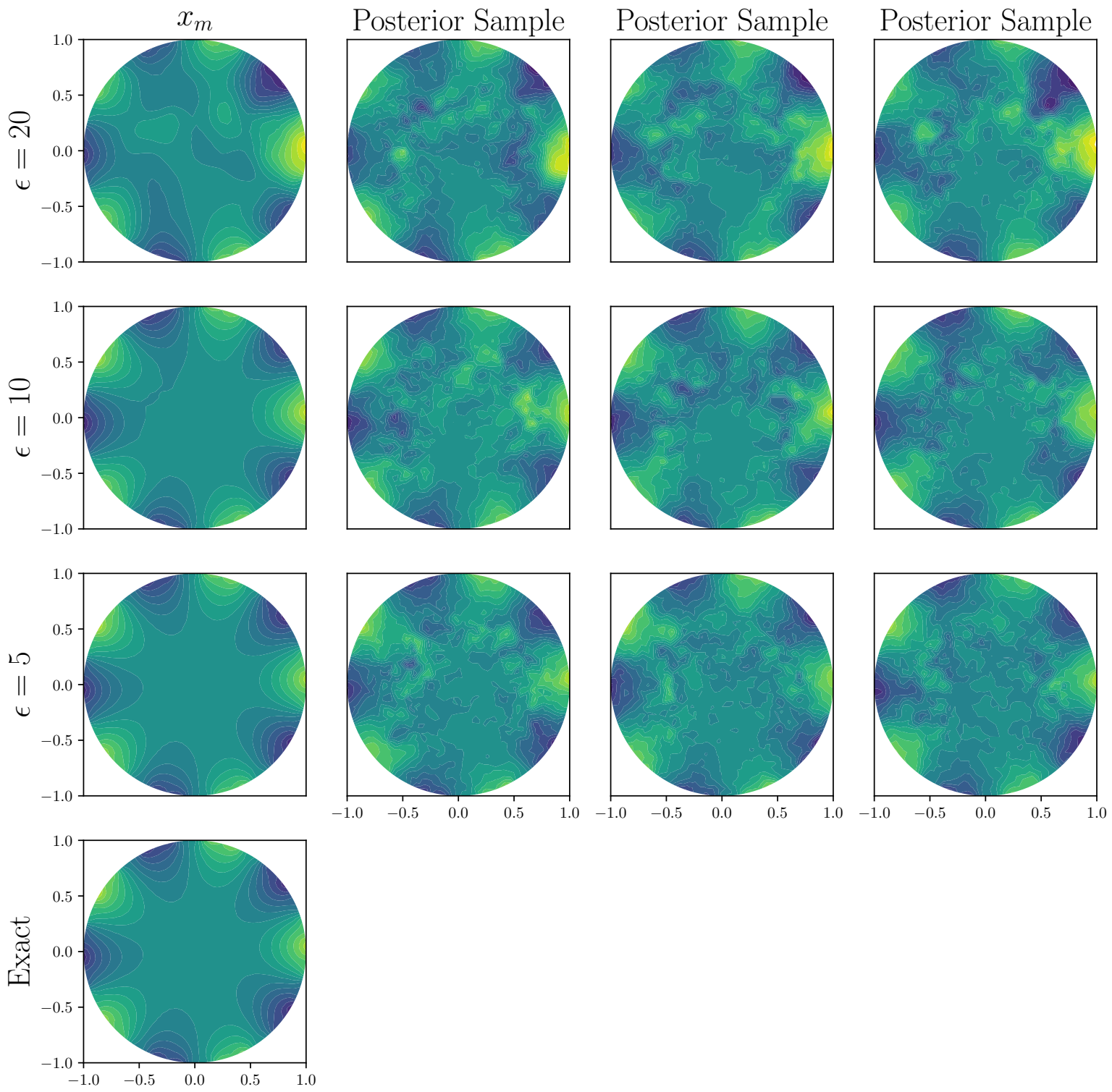

Figure S5: The posterior mean (left column), and samples from the posterior distribution (other columns) produced by BayesCG. The bottom panel represents the actual posterior mean, obtained when the defining linear systems are exactly solved.

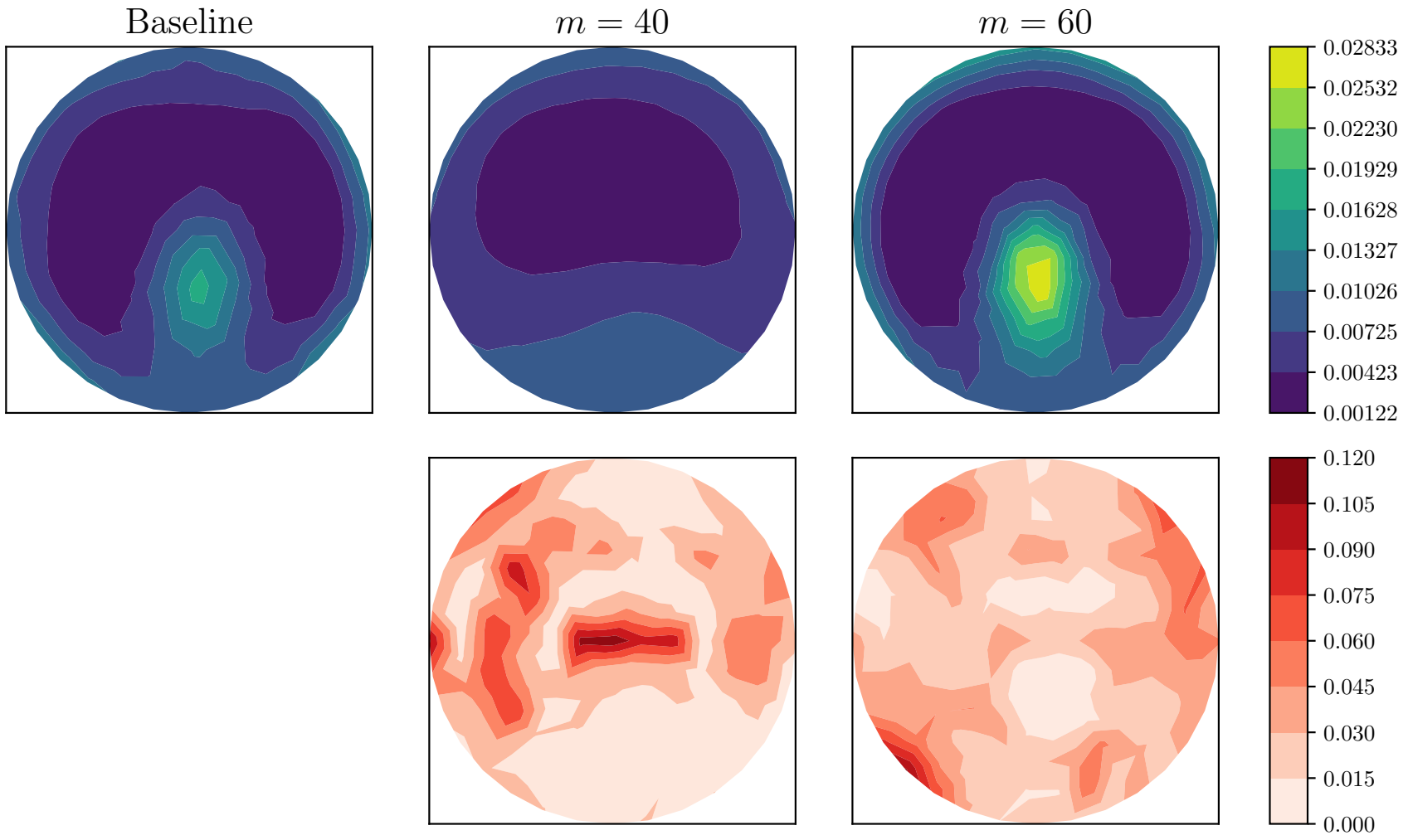

Figure S6: Posterior means (top row) and standard deviations relative to the actual ("baseline") posterior (bottom row) for the EIT problem with a standard CG forward solver, as $m$ is varied.

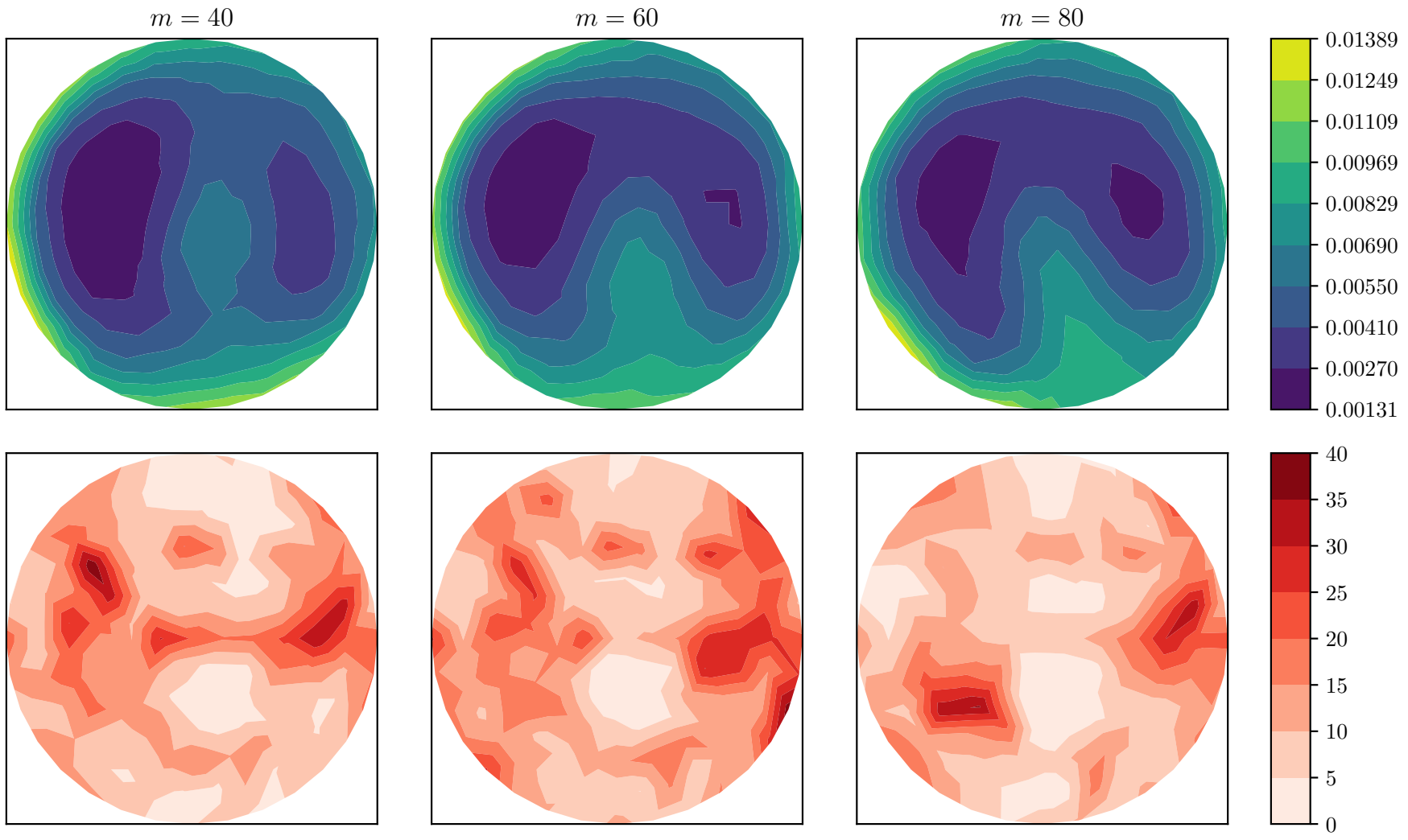

Figure S7: Posterior means (top row) and standard deviations relative to the actual ("baseline") posterior (bottom row) for the EIT problem using a BayesCG forward solver, as $m$ is varied.